\documentstyle[twoside]{article}

\catcode`\@=11
\long\def\@makefntext#1{ 
\protect\noindent \hbox to 3.2pt {\hskip-.9pt
$^{{\eightrm\@thefnmark}}$\hfil}#1\hfill} 

\def\thefootnote{\fnsymbol{footnote}}
 \def\@makefnmark{\hbox to 0pt{$^{\@thefnmark}$\hss}}  

\def\ps@myheadings{\let\@mkboth\@gobbletwo
\def\@oddhead{\hbox{} 
\rightmark\hfil\eightrm\thepage}
\def\@oddfoot{}\def\@evenhead{\eightrm\thepage\hfil 
\leftmark\hbox{}}\def\@evenfoot{}
\def\sectionmark##1{}\def\subsectionmark##1{}}

\oddsidemargin=\evensidemargin
\renewcommand{\thefootnote}{\fnsymbol{footnote}}

\newcounter{sectionc}\newcounter{subsectionc}\newcounter{subsubsectionc}
\renewcommand{\section}[1] {\vspace{12pt}\addtocounter{sectionc}{1}
\setcounter{subsectionc}{0}\setcounter{subsubsectionc}{0}\noindent
	{\tenbf\thesectionc. #1}\par\vspace{5pt}}
\renewcommand{\subsection}[1] {\vspace{12pt}\addtocounter{subsectionc}{1}
	\setcounter{subsubsectionc}{0}\noindent
	{\bf\thesectionc.\thesubsectionc. {\kern1pt \bfit #1}}\par\vspace{5pt}}
\renewcommand{\subsubsection}[1] {\vspace{12pt}\addtocounter{subsubsectionc}{1}
	\noindent{\tenrm\thesectionc.\thesubsectionc.\thesubsubsectionc.
	{\kern1pt \tenit #1}}\par\vspace{5pt}}
\newcommand{\nonumsection}[1] {\vspace{12pt}\noindent{\tenbf #1}
	\par\vspace{5pt}}

\newcounter{appendixc}
\newcounter{subappendixc}[appendixc]
\newcounter{subsubappendixc}[subappendixc]
\renewcommand{\thesubappendixc}{\Alph{appendixc}.\arabic{subappendixc}}
\renewcommand{\thesubsubappendixc}
	{\Alph{appendixc}.\arabic{subappendixc}.\arabic{subsubappendixc}}

\renewcommand{\appendix}[1] {\vspace{12pt}
        \refstepcounter{appendixc}
        \setcounter{figure}{0}
        \setcounter{table}{0}
        \setcounter{lemma}{0}
        \setcounter{theorem}{0}
        \setcounter{corollary}{0}
        \setcounter{definition}{0}
        \setcounter{equation}{0}
        \renewcommand{\thefigure}{\Alph{appendixc}.\arabic{figure}}
        \renewcommand{\thetable}{\Alph{appendixc}.\arabic{table}}
        \renewcommand{\theappendixc}{\Alph{appendixc}}
        \renewcommand{\thelemma}{\Alph{appendixc}.\arabic{lemma}}
        \renewcommand{\thetheorem}{\Alph{appendixc}.\arabic{theorem}}
        \renewcommand{\thedefinition}{\Alph{appendixc}.\arabic{definition}}
        \renewcommand{\thecorollary}{\Alph{appendixc}.\arabic{corollary}}
        \renewcommand{\theequation}{\Alph{appendixc}.\arabic{equation}}
        \noindent{\tenbf Appendix \theappendixc #1}\par\vspace{5pt}}
\newcommand{\subappendix}[1] {\vspace{12pt}
        \refstepcounter{subappendixc}
        \noindent{\bf Appendix \thesubappendixc. {\kern1pt \bfit #1}}
	\par\vspace{5pt}}
\newcommand{\subsubappendix}[1] {\vspace{12pt}
        \refstepcounter{subsubappendixc}
        \noindent{\rm Appendix \thesubsubappendixc. {\kern1pt \tenit #1}}
	\par\vspace{5pt}}

\newcommand{\textlineskip}{\baselineskip=13pt}
\newcommand{\smalllineskip}{\baselineskip=10pt}

\def\eightcirc{
\begin{picture}(0,0)
\put(4.4,1.8){\circle{6.5}}
\end{picture}}
\def\eightcopyright{\eightcirc\kern2.7pt\hbox{\eightrm c}}

\def\abstracts#1#2#3{{
	\centering{\begin{minipage}{4.5in}\baselineskip=10pt\eightrm
	\centerline{ABSTRACT}
	\parindent=0pt #1\par
	\parindent=15pt #2\par
	\parindent=15pt #3
	\end{minipage} }\par}}

\renewenvironment{thebibliography}[1]			
	{\ninerm
	 \baselineskip=11pt				
	 \begin{list}{\arabic{enumi}.}
	{\usecounter{enumi}\setlength{\parsep}{0pt}
	 \setlength{\leftmargin 17pt}{\rightmargin 0pt}	
	 \setlength{\itemsep}{0pt} \settowidth		
	{\labelwidth}{#1.}\sloppy}}{\end{list}}

\newcounter{itemlistc}
\newcounter{romanlistc}
\newcounter{alphlistc}
\newcounter{arabiclistc}

\newcommand{\fcaption}[1]{
        \refstepcounter{figure}
        \setbox\@tempboxa = \hbox{\eightrm Fig.~\thefigure. #1}
        \ifdim \wd\@tempboxa > 5in
           {\begin{center}
        \parbox{5in}{\eightrm \smalllineskip Fig.~\thefigure. #1 }
            \end{center}}
        \else
             {\begin{center}
             {\eightrm Fig.~\thefigure. #1}
              \end{center}}
        \fi}

\newcommand{\tcaption}[1]{
        \refstepcounter{table}
        \setbox\@tempboxa = \hbox{\eightrm Table~\thetable. #1}
        \ifdim \wd\@tempboxa > 5in
           {\begin{center}
        \parbox{5in}{\eightrm\smalllineskip Table~\thetable. #1 }
            \end{center}}
        \else
             {\begin{center}
             {\eightrm Table~\thetable. #1}
              \end{center}}
        \fi}

\def\@citex[#1]#2{\if@filesw\immediate\write\@auxout	
	{\string\citation{#2}}\fi			
\def\@citea{}\@cite{\@for\@citeb:=#2\do			
	{\@citea\def\@citea{,}\@ifundefined		
	{b@\@citeb}{{\bf ?}\@warning
	{Citation `\@citeb' on page \thepage \space undefined}}
	{\csname b@\@citeb\endcsname}}}{#1}}

\newif\if@cghi
\def\cite{\@cghitrue\@ifnextchar [{\@tempswatrue
	\@citex}{\@tempswafalse\@citex[]}}
\def\citelow{\@cghifalse\@ifnextchar [{\@tempswatrue
	\@citex}{\@tempswafalse\@citex[]}}
\def\@cite#1#2{{$\null^{#1}$\if@tempswa\typeout
	{IJCGA warning: optional citation argument
	ignored: `#2'} \fi}}

\def\pmb#1{\setbox0=\hbox{#1}
	\kern-.025em\copy0\kern-\wd0
	\kern.05em\copy0\kern-\wd0
	\kern-.025em\raise.0433em\box0}

\def\fnm#1{$^{\mbox{\scriptsize #1}}$}
\def\fnt#1#2{\footnotetext{\kern-.3em
	{$^{\mbox{\scriptsize #1}}$}{#2}}}

\def\fpage#1{\begingroup
\voffset=.3in
\thispagestyle{empty}\begin{table}[b]\centerline{\footnotesize #1}
	\end{table}\endgroup}

\def\runninghead#1#2{\pagestyle{myheadings}
\markboth{{\eightit{\quad #1}}\hfill}{\hfill{\eightit{#2\quad}}}}

\font\tenbf=cmbx10
\font\tenit=cmti10
\font\tenit=cmti10
\font\bfit=cmbxti10 at 10pt
 1
 1
 1

\font\ninerm=cmr9

\font\eightrm=cmr8
\font\eightit=cmti8

\def\qed{\hbox{${\vcenter{\vbox{                          
   \hrule height 0.4pt\hbox{\vrule width 0.4pt height 6pt
   \kern5pt\vrule width 0.4pt}\hrule height 0.4pt}}}$}}

\runninghead{Fermions, anomaly and unitarity ...}
{R. Guida, K. Konishi and N. Magnoli }

\begin{document}
\normalsize\textlineskip
{\thispagestyle{empty}
\setcounter{page}{1}

\renewcommand{\thefootnote}{\fnsymbol{footnote}} 


\fpage{1}
\rightline{\bf GEF-Th-17/1993}
\rightline{hep-ph/9311219}
\vspace{0.35truein}

\centerline{\bf FERMIONS, ANOMALY AND UNITARITY}
\vspace*{0.035truein}
\centerline{\bf IN HIGH-ENERGY ELECTROWEAK INTERACTIONS}
\vspace{0.37truein}
\centerline{\footnotesize R. GUIDA, K. KONISHI and N. MAGNOLI}
\vspace*{0.015truein}
\centerline{\footnotesize\it  Dipartimento di Fisica, Universit\`a di
Genova, Italy }
\centerline{\eightrm and}
\centerline{\footnotesize\it  INFN, Sezione di Genova}
\vspace*{0.025truein}
\centerline{\footnotesize\it   Via Dodecaneso, 33,  16146 Genova, Italy}
\vspace{0.225truein}


\vspace*{0.21truein}
\abstracts{\noindent We report  the  "state of the art" of the problem  of
$B+L$ violation  in  high-energy  electroweak scatterings.  Results of various
analyses  point toward (though do not prove rigorously yet) the
"half-suppression", i.e., that   the $B+L$ violating
 cross section
remains suppressed at least by the negative exponent of the single
instanton action, at all energies.
Most interesting   techniques developed in this field are reviewed.
Particular attention is paid to unitarity constraints
 on the anomalous
cross section, and to some conceptual problem
involving the use of the optical theorem in the presence of instantons.}{}{}

\vspace*{-3pt}

\textlineskip



\newcommand{\beq}{\begin{equation}}
\newcommand{\eeq}{\end{equation}}
\newcommand{\bea}{\begin{eqnarray}}
\newcommand{\eea}{\end{eqnarray}}
\newcommand{\beas}{\begin{eqnarray*}}
\newcommand{\eeas}{\end{eqnarray*}}
\newcommand{\non}{\nonumber}
\newcommand{\defi}{\stackrel{\rm def}{=}}
\def\de{\partial}
\def\si{\sigma}
\def\sb{{\bar \sigma}}
\def\rn{{\bf R}^n}
\def\r4{{\bf R}^4}
\def\s4{{\bf S}^4}
\def\Tr{\hbox{\rm Tr}}
\def\ker{\hbox{\rm ker}}
\def\dim{\hbox{\rm dim}}
\def\sup{\hbox{\rm sup}}
\def\inf{\hbox{\rm inf}}
\def\re{\hbox{\rm Re}}
\def\im{\hbox{\rm Im}}
\def\infi{\infty}
\def\nrm{\parallel}
\def\nrmi{\parallel_\infty}
\def\teo{\noindent{\bf Theorem}\ }
\def\bra{\langle}
\def\ket{\rangle}
\def\e{x}
\def\sp{E_{sp}}
\def\all{\hbox{\rm all}}
\def\co{$^,$}
\def\daa{$^-$}
\def\dirac{{\cal D}}
\def\dplus{{\cal D_{+}}}
\def\dminus{{\cal D_{-}}}
\def\om{\Omega}
\def\dinv{{\bar D}^{-1}}
\def\et0{\eta^{(a)}_0}
\def\emi{\eta^{(i)}_m}
\def\zema{{\bar \zeta}^{(a)}_m}
\def\etm{\eta^{(a)}_m}
\def\etn{\eta^{(a)}_n}
\def\zet0{{\bar \zeta}^{(i)}_0}
\def\zetn{{\bar \zeta}^{(i)}_n}
\def\zetm{{\bar \zeta}^{(i)}_m}
\def\dainv{({\bar D}^{(a)})^{-1}}
\def\cbar{{\bar C}}
\def\bbar{{\bar B}}
\def\d00{{\bar D}_{00}}
\def\dbar{{\bar D}}
\def\dabar{{\bar D}^{(a)}}
\def\dibar{{\bar D}^{(i)}}
\def\proja{{\bf 1}-|a,0 \rangle\langle a,0|}
\def\proji{{\bf 1}-|i,0\rangle \langlei,0|}
\def\sbar{\bar S}
\def\tt{\tilde T}
\def\st{\tilde S}
\def\sprime{S^{\prime}_{x,y} }
\def\i-a{instanton-anti-instanton}
\def\qi{{\cal Q}_i}
\def\calpi{{\cal P}_i}
\def\qa{{\cal Q}_a}
\def\pa{{\cal P}_a}
\def\calf{{\cal F}}
\def\calg{{\cal G}}
\def\Bvio{\Delta (B+L) \ne 0}

\def\np#1#2#3{ Nucl. Phys. {\bf #1} (#2) #3}
\def\pl#1#2#3{ Phys. Lett. {\bf #1} (#2) #3}
\def\pr#1#2#3{ Phys. Rev. {\bf #1} (#2) #3}
\def\prep#1#2#3{ Phys. Rep. {\bf #1} (#2) #3}
\def\prl#1#2#3{ Phys. Rev. Lett. {\bf #1} (#2) #3}
\def\mpl#1#2#3{ Mod. Phys. Lett. {\bf #1} (#2) #3}
\def\jp#1#2#3{ Journ. of Physics, {\bf #1} (#2) #3}
\def\jmp#1#2#3{J. of Math. Phys., {\bf #1} (#2) #3}

\def \s2{\sigma_2}
\def \S{\hbox{\it S}}
\def\O{\hbox{\it O}}
\def\P{\hbox{\it P}}
\def \bb{\bar b}
\def \ba{\bar a}
\def \half{{1\over 2}}
\def \ak{a_{{\bf k}}}
\def \amk{a_{{\bf -k}}}
\def \bk{b_{{\bf k}}}
\def \bmk{b_{{\bf -k}}}
\def \bak{\bar a_{{\bf k}}}
\def \bamk{\bar a_{{\bf -k}}}
\def \bbmk{\bar b_{{\bf -k}}}
\def \bbk{\bar b_{{\bf k}}}
\def \om {\omega }
\def \Om{\Omega}
\def \omk{\omega _{{\bf k}}}
\def \k{{\bf k}}
\def \dfi{\dot\phi}
\def\const{\hbox{\rm const.} }
\def\t{\tau}
\def\ie{\hbox{i.e.~}}

\def\H{{\cal H}}

\def\q0n{\bra N|q|0\ket }
\def\qmn{\bra N|q^I|M\ket }


\section{Introduction}
We first present  an overview of
 the progress made in the last few years
in the problem of $B+L$ violation in high energy electroweak processes.
  The earlier part of the
development  will then be reviewed, which serves as a
technical  introduction to the subsequent sections.

\subsection{Overview}

In the standard electroweak theory, the baryon (and  lepton-) number
is not strictly conserved\cite{thoft76} as a consequence of the chiral
anomaly\cite{abj}
\begin{equation}
\de_{\mu} J^{\mu}=\frac{g^2}{16\pi ^2}\Tr F_{\mu \nu}{\tilde F}^{\mu \nu}
\end{equation}
where
\beq J^{\mu} \equiv       \bar{ \psi}_L \gamma_\mu  \psi _L  \eeq
is the chiral current for a  lefthanded doublet $\psi_L $ in the theory.
 Due to dynamical, non perturbative effects (such as instantons  or sphalerons)
this leads to physical processes  with the selection rule,
\beq
\label{stanco}
   \Delta (B+L) =(3 + 3) \Delta {\cal N}_{CS};\qquad   \Delta (B-L) =0, \eeq
where ${\cal N}_{CS}$ is the Chern Simons number (see Eq.(\ref{csn})).
For instance, an  instanton leads to
 elementary processes such as
\begin{equation}
q_L +q_L \rightarrow 7\,\overline{q}_L + 3\overline{l}_L + n_{w}W+ n_{z}Z
+n_{h}H.
\label{bvprocess}\end{equation}

The instanton describes a transition between the neighboring classical
vacua\cite{jackiw}\co\cite{gross}.   It is well known that in the
standard electroweak model the barrier
 height corresponds to the
energy of the sphaleron configuration\cite{manton},
\beq  \sp \equiv {\pi \over \alpha} M_W  \sim 10 \,{\rm TeV}, \eeq
(where $\alpha\equiv\alpha_W\sim 1/32$.)
Therefore one  expects  that  at energies much lower than $\sp$
 the  cross sections for such    $\Delta (B+L) \neq 0$ processes are
typically suppressed by the t'Hooft's tunnelling factor,
\beq
e^{-{4\pi\over\alpha}} \sim 10^{-170}
\eeq
utterly  too small for  them to be observable.

Nevertheless,  in at least two
situations, at high temperatures\cite{hight}
 and at very high fermion densities\cite{highd},
baryon number violation is believed to proceed
 unsuppressed. The interest and importance for  such a
possibility is mainly related to the problem of cosmological baryon
number generation in the standard $SU(3) \times SU(2) \times U(1)$
theory of fundamental interactions. For a recent review, see e.g.
Shaposhnikov\cite{shap}.

Whether or not the $\Delta(B+L) \neq 0$ events will become observable
at high energy  {\it scattering processes} (at the energy available at
SSC or LHC), has been the issue of an active debate for a few years now.
{}From experimental physics point of view, the prospect of observing roughly
isotropic production of a large number $ n=O({1 \over \alpha})$ of W or Z
bosons is quite an exciting one\cite{moriond}.
{}From the theoretical point of view, the problem
involves some of the most subtle aspects of non-Abelian gauge theories,
such as  chiral anomaly,  degenerate classical  vacua   and
 quantum tunnelling among them,
 large order breakdown of  perturbation theory,  compatibility of
semiclassical expansions with  unitarity, and so on.

It is the purpose of the
present paper to review the latest developments in this field of research
and assess our general understanding.
Earlier works have been fully reviewed by Mattis\cite{mattis}. See also
the proceedings  of the Santa Fe workshop\cite{sfe}.
For a more  recent review, see Tinyakov\cite{tinyakov}.

\smallskip

Historically, after first suggestions\cite{aoya87}\co\cite{mclerd87},
 the semi-classical estimate of the total cross section
with $\Delta(B+L) \neq 0$ was shown\cite{ring90}\co\cite{esp90}
 to grow exponentially with energy
(at low energies), stimulating further
works\cite{mcler90}$^-$\cite{vol91d}. It was noted
 immediately\cite{zak90}
that the  growth could not continue indefinitely as the computed cross section
 violated the unitarity limit above  the sphaleron mass energy
$\sp $.
Quantum corrections around the instanton were
studied\cite{zak90}$^-$\cite{porrati}
 and it was shown that
 all tree type corrections involving the final states (so
called soft-soft corrections) contributed to the nontrivial, exponential energy
dependence. In particular, it was argued\cite{mat90}
 that these corrections could be
expressed in the form,
\beq\label{holy}
\sigma_{\Delta(B+L) \neq 0}=e^{ {4\pi\over\alpha} F(\e )}
\eeq
where the function $F(\e )$ (called sometimes "Holy Grail" function) has an
expansion in $ \e \equiv E/\sp \ll 1$.

 Two distinct approaches, the R term method\cite{krt}\co\cite{esp93}
 and the  valley
method\cite{kr91b}$^-$\cite{bal91}, were developed to compute the anomalous
cross section, and used in recent perturbative
calculations\cite{dp91}\daa\cite{silvestrov}.
Subsequently
the equivalence of the two methods as regards  the final state corrections,
 has been argued to hold to all orders of perturbations around the
instanton\cite{mat91b}.
The developments which followed, to be discussed
in Sections
2 and 4, make crucial use of these methods.

A  non perturbative model,  utilizing the valley field and the optical
theorem, was presented by Khoze and Ringwald\cite{kr91}, with the aim of
resumming  all final tree corrections.

Other kinds of corrections involving the initial, hard particles
(the so called hard-hard and hard-soft corrections)
 were then studied\cite{mu90}\daa\cite{mcler91}, following the
impetus provided by  the works of Mueller.
The results suggest the exponentiation of
loop contributions to all orders,
 however, in the direction of suppressing the
$\Delta (B+L)$ cross section.

Such an  exponentiation
supports strongly the idea that the quantum corrections involving the initial
hard particles can  also be included in a modified semi-classical
approximation. Led by this thought several different approaches have
been proposed.
Works reviewed in Section 2 may indeed   be called
"Search for the new semi-classical field", or perhaps, more
 poetically, "Search for
the Holy Grail."

Mueller\cite{mu92} and independently Mc Lerran et
al.\cite{mcler92}, propose a
classical equation of motion for the fields, with the source term to take
 account of  the initial energetic particles. They show that to
lowest orders their solution automatically reproduces the quantum loop
corrections found earlier by direct calculations. However, as pointed out
by these authors themselves, the necessity of using Minkowskian time
or complex time (and in general complex fields) and the consequent
complexity of the field equation involved, seem to make the task
rather a formidable one.

An alternative approach was proposed by Rubakov and
Tinyakov\cite{tin92}\daa\cite{kya92b}.
They propose to study,
instead of the original $2 \rightarrow \all$ type cross sections, the
processes involving initial many-body states, $n \rightarrow \all, n=
{ \nu \over g^2 }$, in the limit $g^2 \rightarrow 0, $   with fixed $\nu
$ and fixed   total initial energy E. The idea is that once
 a semi-classical approximation for such many-body processes has been
established, one can then take the limit $\nu \rightarrow 0$, hopefully
recovering the answer for the $2 \rightarrow \all$ process.
The coherent state formalism first introduced in the $ \Delta (B+L) \neq 0$
problem by Khlebnikov and others\cite{krt} plays a powerful
 role in this approach.
Latest work\cite{mu92b} seems  to show that the
 above limit is indeed a smooth one.

It is however not clear at the moment whether these approaches can lead to
 a new truly  semi-classical approximation and what the eventual
 answer might be.

Perhaps the clearest physical picture of what happens in high-energy
$\Bvio$ processes, has been given by the study of quantum-mechanical
analogue problems\cite{girard}\daa\cite{bac92}.
  There are two competing factors which determine  the
energy dependence of the cross section\cite{banks90}.
 One, the semi-classical
tunneling factor, rapidly grows with energy, and the suppression is
altogether lifted when the sphaleron energy is reached.  Another factor
is the probability amplitude - mismatch - between the most favorable
state (for the purpose of tunnelling) and the initial two-particle state.
And this factor (an analogue of Landau's semiclassical matrix elements)
gets strongly damped as the energy increases. As a consequence
the full amplitude never
gets large, leading  to the "half-suppression" result (see below).
 Diakonov and Petrov\cite{Diakonov93} recently applied to the
problem a generalization\cite{pit} of such a WKB
approximation to field theory,  finding some indication for
  the suppression of the cross section for the
isotropic production of many $W$'s,  even at very high energies: $E\gg \sp$.

Quite parallel to the developments  mentioned above, several
 arguments,  essentially all taking unitarity costraints
into account, were presented\cite{zak91b}\daa\cite{kkr92b},
  which   suggested that the $\Delta (B+L) \neq 0$
cross section remains suppressed {\it at least} by
\beq
e^{- {2\pi \over \alpha}}
\eeq
(so-called "half suppression").  In spite of their
apparent differences, these arguments are actually closely  interrelated,
and depend only on the S-wave unitarity and the dominance of multiparticle
production events.

The whole question was  analysed  from a somewhat different angle by the
present authors\cite{guida}\co\cite{Nico}.
 The  approach using the optical theorem and the valley field
 to compute  the
total anomalous cross section\cite{kr91},
 takes  unitarity into account  automatically,  and
 furthermore  seems to enable one to compute the Holy Grail function
 up to the sphaleron energy.  However,
  a new
kind of problem arises (which may be
termed the "unitarity puzzle").
  Namely, how can one extract the part of the imaginary
part of the elastic amplitude, that corresponds to the $\Delta (B+L) \neq
0$ intermediate states? In other words, how is unitarity satisfied in
the presence of topologically nontrivial effects such as instantons ?
A partial answer  was  given by the equivalence proof by Arnold and
Mattis\cite{mat91b}; however the crucial issue concerns
the initial particles, see Section 5.  Or, when does an
 "instanton-anti-instanton"
 type configuration
cease to be topologically nontrivial? To clarify these issues  requires
a  detailed study of the behavior of chiral fermions in a background of
instanton anti-instanton type (hence the title of this review!).  The
 result of this investigation leads us once more
to the above mentioned
 "half-suppression" of the $ \Delta (B+L) \neq 0$ cross section.

\bigskip
The rest of the paper is  organized as follows.  In subsections
1.2, 1.3 and 1.4 the earlier developments, including  the original instanton
calculation,  the $R$-term method  and the valley method,  are
reviewed, which also serve for fixing the convention and notation for
later sections.  The reader already involved in the research
in this field may
well skip this section; for others this section should  provide an
appropriate technical introduction.

The Section 2 contains a somewhat detailed review of  more recent,
new  semi-classical approaches
to the Holy Grail function. The discussion is divided into two parts, the
first (subsection 2.1) dealing with attempts to take into account the
effects of the
initial high energy particles into a semi-classical equation, the second
(subsection 2.2) being mainly
concerned  with the multi-particle approach of Rubakov and Tinyakov.

 In Section 3  we review
those studies based
on quantum mechanical analogue problems, which appear to provide  an
intuitive understanding of the whole problem.

In Section 4  various analyses,  leading to  unitarity bounds on the
baryon number violation in high energy  electroweak scatterings,
 are reviewed. We
first discuss a simple, multi-instanton unitarization picture and the general
unitarity bound following from the S-wave dominance (subsection 4.1).
The results are corroborated by  an explicit resummation of  multi-instanton
contributions  in subsection 4.2. The physical difference between
 the high {\it temperature}
or high {\it density} transitions and the high {\it energy} $B+L$ violation
is briefly mentioned in subsection 4.3.

The calculation of the anomalous cross section via optical theorem, done
with the valley method, is critically analysed in Section 5.  After the
discussion of the unitarity puzzle (subsection 5.1), the fermion Green
function in the background of the valley field is studied and the
anomalous part of the forward elastic amplitude identified (subsection 5.2).
The melting of the instanton anti-instanton pair, and the ensuing
transition to purely perturbative amplitude, is discussed in
subsections 5.3 (through the study of the Chern-Simons number) and
in 5.4 (in which  the fermion level
crossing in the valley is analysed).   The implication of these results
to high energy electroweak processes  is  summarized in subsection 5.5.

We conclude (Section 6) by discussing    a unifying and consistent
 picture which seems to emerge
through different types of analyses reviewed here.



\subsection{The original instanton calculation}

In this section we recall briefly  the
 calculation of  Ringwald\cite{ring90}
and Espinosa\cite{esp90}  for the cross
section of the process (\ref{bvprocess}).
See Mattis\cite{mattis} for more details.
In order to keep formulas as simple as possible we shall take only
 fermions and
W bosons as external particles.
Let us consider  the Euclidean $n+ N_F $ point
Green function ($N_F$ being the number of the lefthanded fermion doublets),
$$ G(x_1 ,\ldots x_{n}, y_1, \ldots y_{N_F} )=  $$
\beq \int dA \,d\psi\, d{\bar \psi}\, d\phi \,
 A_{\mu _1}\!(x_1 )\ldots
A_{\mu _{n}}\!(x_{n})\, \psi (y_1\!) \ldots \psi (y_{N_F}\!) \,
 e^{ -S(\phi , A_{\mu }, \psi , {\bar \psi } )}.
\label{green}
\eeq
This Green function gets contributions only from gauge fields with unit
topological number $Q=1$, where
\beq
Q\equiv
\int d^4x {g^2\over 16 \pi^2}
\hbox{\rm Tr }F_{\mu \nu}
 {\tilde F}_{\mu \nu} = {\cal N}_{CS}(-\infty)-{\cal N}_{CS}(\infty),
\eeq
(${\cal N}_{CS}$ is the Chern Simons number, see  Eq.(\ref{csn})).
In the semi-classical approximation one needs
appropriate
 classical solutions  with a finite action.
In the electroweak theory one can
show that no classical solutions  exist
  in the presence of the Higgs vacuum expectation value.
    Indeed if one scales
  \beq  \phi (x) \to \phi (ax);\qquad A_{\mu}(x) \to a A_\mu(ax), \eeq
(which is
  a particular kind of variation),
  and considers the action  as a function of $a$,  one  sees
  that no minimum exists for $a \ne 0$, for any functions $\phi $ and $A$.
    On the other hand the existence
  of a classical solution would require the action to be  a minimum under
  {\it all \,} variations around it.

One is forced to look for a constrained instanton solution\cite{affleck}.
 Although the  solution  is not known  in  a closed
form, one knows asymptotic forms when the distance
from the center of the instanton  $x_i$ is much less or greater
than the size $\rho$ of the instanton.  Very near the instanton center it is
given by
\bea  A_\mu ^{(inst)}(x) \simeq
-{i\over g}U (\sigma_{\mu} {\bar \sigma_{\nu}}
-\delta_{\mu \nu} ) U^{\dagger}\,
{(x-x_i)_{\nu} \rho^2 \over (x-x_i)^2 ((x-x_i)^2 +\rho^2) },  \non \\
\phi ^{(inst)}(x) \simeq  U
\left( \begin{array}{c}
0 \\
{v/ \sqrt{2}} \end{array} \right) \left({(x-x_i)^2 \over ((x-x_i)^2
+\rho^2)}\right)^{1/2}, \qquad   {(x-x_i)^2 \over \rho ^2 } \ll 1;
   \label{instanton}\eea
where $\si_\mu\equiv (i, \vec{\si}),$   $\sb_\mu\equiv (\si_\mu)^{\dagger}$
and $U$ is the global $SU(2)$ rotation.
Far away from the center  $ A_\mu ^{(inst)}(x)$ is proportional to a
massive boson propagator; $\phi ^{(inst)}(x)$ approaches a constant,
$\phi ^{(inst)}(x) \simeq  U
\left( \begin{array}{c}
0 \\
{v/ \sqrt{2}} \end{array} \right). $

The leading order approximation consists in keeping only  up to the quadratic
part of  the
fluctuations in the action and in substituting the fields by the classical
ones  in
the pre-exponential factors. For fermions the "classical solutions" are
the zero modes of the Dirac operator in the classical background, which
behave as a free massless propagator asymptotically.
The gauge  field satisfies the massive
free field equation far from the instanton center:
 its Fourier transform displays a pole at $k^2 = M_{W}^2$.
Through the LSZ procedure, one  gets  the residue of   this
pole equal to
\beq
{\cal R}_\mu ^a(\rho,{\bf k})=
{4\pi ^2 i \over g} \rho ^2 {\bar \eta } ^a _{\mu \nu} k_{\nu},
\label{residue}\eeq
where ${\bar \eta }$ is the usual t'Hooft symbol\cite{thoft76}.
Each  fermion doublet contributes with a  factor $\sim e^{ikx_i} \rho $,
coming from the LSZ amputation of  the zero mode,
\beq  \psi _0 (x) {\sim}{\rho \over  (x-x_i)^3}\qquad
 {\hbox{\rm for}}\,\,\,{x \to \infty} .\eeq
The action evaluated at the classical field, is equal to\cite{affleck}
\beq
 S_{c}={2\pi \over \alpha}  + \pi ^2 \rho ^2 v^2
+O(\rho^4 v^2 M_H^2 \log (M_H \rho )).
\eeq
Putting  all these pieces together, and integrating over the collective
coordinates (the instanton position, size and the  isospin orientation
$U$)
  one finds
\beq
A_{2 \rightarrow n} \sim n!({1 \over v^2 g})^{n} \exp(-S_{inst}) \mid \k _1
\mid ....\mid \k _{n+2} \mid
\eeq
where $S_{inst} \equiv 2\pi / \alpha $ and $n!$ comes from the integration
over the size,
\beq \int_0^\infi d\rho \, \rho ^{2n} e^{-\pi ^2 \rho ^2 v^2}. \eeq

This gives the S matrix element at fixed $n$. The total cross section
is found  by squaring it and summing over the final states. To understand
qualitatively its behavior\cite{mattis}, recall that the
 relativistic phase space goes as $ {\bar E}^{2n}/n!$; the squared amplitude
as $(n!)^2 {\bar E}^{2n}$, where  ${\bar E} \sim E/n$ is the average
momentum. For large $n$ these  together give rise to an "exponential"
series $ \sum_n ( E^{4/3})^{3n}/(3n)!.$
A more careful estimation\cite{zak90}\co\cite{porrati}\co\cite{krt}
  leads to
\bea
\sigma _{2 \rightarrow \all}  &\propto&  {1\over s } \sum_n
{1\over (n!)^3} ({3s^2\over 8g^2\pi ^2v^4})^n e^{-2S_{inst}} \non \\
&\sim &  {1\over s}e^{{4\pi \over \alpha}(-1+
{9 \over 8}({E\over E_0})^{{4 \over 3}})},
\label{growth}\eea
where $E_0 \equiv \sqrt{6} \pi  M_W/ \alpha.$

The cross-section grows exponentially with energy.  If
such a growth should  continue up to and above  the sphaleron energy,
\beq
\sp \equiv{\pi \over \alpha} M_W, \eeq
the exponential suppression factor would be
 compensated altogether.  Actually, the approximations leading to
 Eq.(\ref{growth}) are valid for energies much less than the sphaleron
energy  as will be  seen below.

   A simple calculation shows that the cross section
is  dominated by  the production  of  W bosons,
with the  average number
\beq
n \sim {1 \over \alpha } x^{4/3},
\label{n}\eeq
where the definition
\beq x \equiv {E\over \sp} \eeq
will be used throughout this review.   $x$ here must be small for these
approximate estimates to be valid, but should not be too small either,
so that the average multiplicity (\ref{n}) is sufficiently larger than unity.
Eq.(\ref{n}) then implies  that the mean energy of the final particles
is  $E/n \sim  M_W x^{-1/3}$,  showing  that
 the final state particles carry soft momenta, while those in the
initial ones  hard (large) momenta.

The inclusion of the Higgs particles does not change  these
features in any essential manner.

Eq.({\ref{growth}), if extrapolated above the sphaleron energy, violates
the
unitarity limit.  It must  therefore be substantially
 corrected before
 the initial
energy reaches $E_{sp}$.  Does the $\Bvio$ cross section
nonetheless grow up to the point of saturating
 the unitarity limit, i.e.,  to a geometrical size, $\sim 1/s$?
  We are
 interested here in those corrections which
exponentiate and modify the energy dependence in the exponent of the
result, \ie , those  which
contribute to
\beq F(x) = \lim_{g \rightarrow 0} \log \sigma _{2 \to \all}. \eeq
$F$   is called  the Holy Grail function.

The corrections to the leading instanton result,  are often
classified into three groups (see Fig. 1):  "soft-soft",
"hard-hard" and "soft-hard" corrections.  The first type involves only the
final
particles, the second one the initial particles and the third one
both the initial and final particles. Also,  contributions coming
from other classical solutions with unit
topological number such as multi-instanton configurations, could
 be important (see Section 4).

All the tree
 soft-soft
corrections have been shown\cite{mat90}  to exponentiate and  give
rise to $1/\alpha $  terms in the exponent.\fnm{a}\fnt{a}{For $x =O(1)$
 the naive perturbation theory
around the instanton would give  an increasingly
divergent  series
of type $ \sim ({1\over \alpha })^n $ which should  necessarily
 be summed to all orders, i.e., nontrivial
corrections to the Holy Grail function.} In order to compute
them, two distinct methods have been
used:
 the valley method and the "$R$-term"
method, which will be reviewed briefly in the next subsections.

\subsection{The $R$-term method}

In the $R$-term method, developed by Khlebnikov, Tinyakov and
Rubakov\cite{krt},
 the summation over the final states  is taken into account  as a  sort of
correction to the action.
This could be   convenient for an application of the steepest descent
method\cite{mat91}.
We follow here an approach to the method
developed by Mc Lerran\cite{mattis}, which is simpler
then the original one\cite{krt}\co\cite{tinyakov}.
The cross section is expressed as a double path integral,
which within the perturbation theory is nothing but
a compact way of implementing the standard  Cutkovsky's rules.

Consider a theory with a scalar field $\phi$ with
Minkowskian action
\beq
S=\int d^4x \,\{ \half(\partial \phi)^2 -\half m^2 \phi ^2 -V(\phi ) \}
\eeq
A Green's function (in momentum space) containing two initial particles
 and n final particles
can be expressed as a path integral
\beq
G(p_1 ,p_2 ,k_1 ,...k_n )=\int d\phi \,\phi (p_1 )^{*}\phi (p_2 )
^{*} \phi (k_1 )....\phi (k_n ) \, e^{iS(\phi ) }.
\eeq
To get the total cross section one applies the LSZ procedure to the
above, squares it (by doubling the fields) and sums over all
 possible final states. These
operations are compactly expressed\cite{mattis}  by  the introduction of
\beq
 R(k )\equiv \lim_{k^2 \rightarrow m^2} (k^2 - m^2)^2 \phi (k)
\phi^{*\prime} (k).
\eeq
$\sigma _{tot}$ can then be written as a double path integral:
\beq
\sigma _{tot} = {1 \over F} \int d\phi \int d\phi ^{\prime}
 R^{*} (p_1 )  R^{*} (p_2 )\,
e^{iS(\phi ) -iS(\phi ^{\prime} )
+\int { d^4 k \over (2\pi )^3}
\delta (k^2 - m^2)\, \theta (k_0 )  R(k_0 )},
\label{rsigma}\eeq
where F is the flux factor, $F\sim s$. Note that, thanks
 to the statistical factor
$1/n!$ for identical particles, the factors associated with
 the final particles
nicely exponentiate upon  summation over $n$, and give  rise to the
third term in the exponent, the so-called $R$ - term.

The $R$-term method is particularly powerful,
when applied to the computation of the corrections to the leading
semi-classical result, Eq.(\ref{growth}).  We shall  here content ourselves
however to observe that
 to lowest order Eq.(\ref{rsigma}) reproduces Eq.(\ref{growth}).

In this approximation one splits the fields into the classical and
 quantum parts,
\beq
\phi (x)= \phi ^{(inst)} (x;\alpha) +\phi^q (x),
\label{split}\eeq
where $\alpha$ indicates the ensemble of the collective coordinates, and
substitutes the classical solution into the $R$ - term, as well as in the
ordinary actions  and in the pre-exponents in (\ref{rsigma}).
In the case of the standard model, the leading contribution to the exponent
comes from $W$ bosons and is represented in Fig. 2.
 Using Eq.(\ref{residue}) and by
carefully continuing from Minkowskian   to Euclidean space, one gets
\beq
\sigma _{tot} \propto
\sum_{\bf n} |<{\bf n}| A | {\bf p},-{\bf p} >|^2 =
 e^{-16 \pi^2 / g^2} \int dt \,d{\bf x} \, d\rho \, d\rho^{\prime}\,
C\,e^W,
\label{Rformula}\eeq
where $(t, {\bf x}) $ is the Minkowski continued difference of the instanton
positions, $x_i - x_i^\prime$ and $ C$ contains the functional
determinant over the nonzero modes, the Jacobian associated to the
introduction of the collective coordinates as well as
 factors from  fermion zero modes\cite{krt}.

 The crucial exponent $W$ is given by
\beq
W=-iEt-\pi^2v^2 (\rho^2 + \rho^{\prime 2}) + {\pi \over g^2}\rho^2
\rho^{\prime 2} \int {d{\bf k} \over \omega_k}  e^{ikx}
(3\omega_k^2 + {\bf k}^2),
\label{expW}\eeq
where
$\omega_k \equiv \sqrt {{\bf k}^2}.$

Integration over ${\bf x}$ by the saddle point method
sets $ {\bf x} = 0$, which leads to
 \beq
W=-iEt-\pi^2v^2 (\rho^2 + \rho^{\prime 2}) + {96\pi^2
\rho^2 \rho^{\prime 2} \over g^2 t^4}.
\label{expW2}\eeq

The last integration over $t$ and $\rho, \rho^{\prime}$  by the
saddle point approximation yields the well known result, Eq.(\ref{growth}).

Higher-order  corrections can be systematically taken into account by
using the expansion Eq.(\ref{split}) in Eq.(\ref{rsigma}).  The resulting
series in the Holy Grail function turns out to be a power series in
$x^{2/3}$ (see Eq.(\ref{holypert}) below.)  It means that in order to get a
significant result for the energy of the order of $\sp$ ($x \sim 1$),
 one must be able to resum all the R-term
contributions corresponding to soft tree graphs, or equivalently
to solve the complete classical equations,
including also the R-term.

It should be emphasized  that the $R$-term  method  (and the $R$-term
corrected classical equation)
only makes sense in Minkowski space.  For instance, by "instantons" one
really means their  analytic continuation to Minkowski space.
Also, a new solution of $R$-term corrected classical equation
with  an $O(3)$ symmetry ("distorted instantons") can be found (though
only in the
linearized approximation)\cite{mat91}, but  it  turns out to be impossible
to continue
it to Euclidean space.  As a result  it is a nontrivial task
 to  get a nonperturbative
control of
 (even) the soft-soft part of  the  Holy Grail function (which we shall call
$F^{ss}(x)$ to distinguish it from full $F(x)$).

The $R$-term method has been generalized by
 Espinosa\cite{esp93}, so as to take account of   the effects of fermion
 pair production.   His
 calculation shows that the inclusion of the fermionic $R$-term does not
 modify  the result to exponential accuracy: it does not affect  the
 Holy Grail function.

\subsection{The valley method}

An alternative approach to the calculation of the
total cross section is based on the optical
theorem\cite{zak90}\co\cite{porrati}
which  relates  it to the imaginary part of the forward elastic amplitude.
However since  we are here interested only in the inclusive $\Bvio$ cross
section, not really the total cross section,
 a highly nontrivial problem
 arises in extracting the "anomalous piece" from the full
imaginary part of the forward elastic amplitude. This issue will be discussed
extensively in Section 5.

$\sigma _{\Bvio}$ is given by
\beq \sigma _{\Bvio} = {1 \over s} {\it Anom}\, {\it Im}\,  {\it LSZ}\,
{\it Wick} \,\int dA\, d\psi \,
d{\bar \psi }\, \psi \psi {\bar \psi} {\bar \psi}
\, e^{-S(A) - \int {\bar \psi }\bar{D}\psi }, \label{optical}\eeq
where with ${\it Anom},\, {\it Im}, \, {\it LSZ},\,
{\it Wick} $ we indicated symbolically the operations of extracting the
{\it anomalous} piece of the  {\it imaginary part} of the on-shell
amplitude ({\it  LSZ  procedure}), from the four point function {\it Wick
continued} to Minkowskian space.
One wishes to evaluate Eq.(\ref{optical}) by a semi-classical approximation.
Since one deals here with an elastic amplitude,  the relevant  gauge
background
must belong to the trivial sector, with zero topological number.   At the
same time, however, it must describe nonperturbative effects of
producing $\Bvio$ intermediate states: it must be  topologically nontrivial
 locally, as  e.g., the instanton anti-instanton pair.
 As is well known from
the example of a  quantum mechanical double well,  a simple sum of
instanton and anti-instanton (at finite distances) is not an adequate
background, since the expansion around it produces a large linear term in
the "quantum" fluctuation, because is not a
solution of field equations.  The effort to minimize the latter term by a shift
of the field introduces automatically an effective interaction between
the instanton and anti-instanton.

The valley (or streamline) method  provides a way to take such
interactions into account systematically. The valley
 method\cite{bal86}\daa\cite{yung}
 is a
generalization of the standard Faddeev-Popov procedure of treating zero
modes in quantum mechanics and in quantum field theory, to the case of
quasi zero modes.   If the action has a valley-like shape in the functional
space, \ie if its value slowly changes along the bottom
 of the valley (streamline),  one must
first perform  Gaussian integrations in directions orthogonal  to the
streamline, then integrate  the result along the streamline.

The valley
trajectory (streamline) $ \phi _\alpha $ is a solution of the
equation,\fnm{b}\fnt{b}{There is another possible way to define the
valley trajectory\protect\cite{aoya92} which is claimed to have some
advantage over Eq.(\ref{valeq}). }
\beq w(x, \alpha ){\de \phi _{\alpha }(x)\over \de \alpha } =
{\delta S \over \delta \phi} |_{\phi =\phi _\alpha
}\label{valeq}\eeq ($w $ is an arbitrary weight function),
$\alpha $ parametrizing the  bottom of the valley.
The generating functional $Z$ in leading approximation
is
\beq
 Z \sim \int d\alpha \, \Vert {\de \phi _\alpha \over \de \alpha} \Vert ^2
\,e^{-S(\phi _\alpha )} \int d\phi \,\delta \left(\int d^4x\,
(\phi -\phi
_\alpha ){\de \phi  \over \de \alpha} w \right) \,
e^{- \int(\phi  - \phi _\alpha )\Box _\alpha (\phi  - \phi _\alpha )}.
 \eeq
This would be the same as the standard functional integration
with a zero mode, were
it not for the nontrivial integration over $\alpha $.

In pure Yang-Mills theory in four dimensions, the valley equation was
solved\cite{yung}\daa\cite{kr91c}
by using an Ansatz,
 $A_\mu =   (2/g) {\bar \sigma }_{\mu \nu} ( x_\nu /x^2) s(x^2). $
 The problem  is then reduced to that of a
simple one-dimensional quantum mechanical double well,  for which the
solution to Eq.(\ref{valeq})
is  known.
The valley trajectory $A_{\mu}^{(valley)}$  found this way
(after a particular conformal transformation of the original
ansatz) is
\bea
A_{\mu}^{(valley)} &=&-{i\over g}(\sigma_{\mu} {\bar \sigma_{\nu}}
-\delta_{\mu \nu} )\,[\, {(x-x_a)_{\nu} \over (x-x_a)^2 +\rho^2 }   \non \\
& &
+
{(x-x_i)_{\nu} \rho^2 \over (x-x_i)^2 ((x-x_i)^2 +\rho^2) }
\non \\ & & + {(x-x_i +y)_{\nu} \over
(x-x_i + y )^2  }  -  {(x-x_i)_{\nu} \over
(x-x_i)^2  }\,]
\label{valley} \eea
We defined
\bea
 y^{\mu} &\equiv& -R^{\mu}/(z-1); \non\\
R^{\mu}&\equiv& (x_i - x_a)^{\mu};\non\\
 z &\equiv& (R^2 + 2 \rho^2 + \sqrt{R^4 + 4\rho^2 R^2}) /2\rho^2,
\label{1_7}\eea
where $x_i^{\mu}$ and $x_a^{\mu}$ are the centers of the instanton and
 anti-instanton, $\rho$ is their (common) size.
As is seen from  Eq.(\ref{valley}) $A_{\mu}^{(valley)}$ interpolates
between two solutions of the classical Yang-Mills equation:  the simple sum
of instanton and anti-instanton at infinite separation ($R=\infty$) and a
gauge-equivalent of $A_\mu = 0$ (at $R=0$).

For simplicity of writing,  the size of the instanton and  that of the
anti-instanton will be  taken to be equal here; no generality  is
however  lost since the
saddle point equations for  $\rho ^a$ and $\rho ^i$ set them equal
anyway, in the problem one is interested  here.

The action of the valley  is given\cite{verb}\co\cite{kr91c}
by\fnm{c}\fnt{c}{The fact that the action depends on the
valley parameters only through $z$ is a reflection of the conformal
invariance of the classical Yang-Mills equation.}

 \beq
S(A_{\mu}^{(valley)}) =
{48 \pi ^2 \over g^2}\left[ {6z^2-14 \over (z-1/z)^2} -{17
\over 3}-\log z\left({(z-5/z)(z+1/z)^2 \over (z-1/z)^3}-1\right)\right]
\label{valac},\eeq
which asymptotically (as $R \to \infty$) behaves as
\beq
  {4\pi \over \alpha}  -  {24 \pi \over \alpha }\,{\rho ^4 \over R^4}
+ O({1 \over R^6}). \label{valacas}
\eeq
The first term is twice the instanton action, while the
 second one represents the attractive interaction between the
instanton pair.

The behavior of the valley action as a function of $R/\rho $ is plotted
in Fig. 3.

In the case of the Weinberg-Salam theory,  the solution of the valley
equation is not known.  At energies much lower than the sphaleron mass,
however, one can justify\cite{zak90}\co\cite{porrati}\co\cite{kr91}the use
of Eq.(\ref{valley}), Eq.(\ref{valacas}) in Eq.(\ref{optical}).
Furthermore the fermion fields  $\psi$'s or ${\bar \psi}$'s
 in Eq.(\ref{optical}) may be replaced by  the standard zero modes
 $\psi_0^{(a)}$'s (in the anti-instanton background) or
 ${\bar \psi}_0^{(i)}$'s (in the instanton background),
respectively, see subsection 5.2.
In this manner
the result
Eq.(\ref{growth}), is reproduced by the valley method, confirming  once
more the
exponential growth of $\sigma _{\Bvio} $ at low energies.

The equivalence between the $R$-term method and the valley
approach has been derived {\it perturbatively} to all orders
in $x$ (or equivalently in ${\rho \over R}$)  in a rather
formal way\cite{mat91b};  their  proof however neglects
incoming particles (see Section 5)
 and also skips the problem of weight dependence
of the result (see Section 2.1).

\smallskip

Without knowing the solution of the complete
 valley equation, the utility of the
valley method in the electroweak theory is  limited, unfortunately.
  Nonetheless,
Khoze and Ringwald\cite{kr91} attempted a nonperturbative calculation of
the Holy Grail function by doing the following simplifications or
assumptions (justified
or not!).  They
{\it (i)} just add to the  valley action Eq.(\ref{valac})
  the Higgs contribution
$-2\pi ^2\rho ^2 v^2$; \,\,
{\it (ii)} substitute  the fermion fields by
the standard fermion zero modes  $\psi_0^{(a)}$'s and  ${\bar
\psi}_0^{(i)}$'s;\,\, and
{\it (iii)} evaluate the resulting integrations
\beq \sigma =Im \int dR\, d\rho \,\exp(ER-2\pi ^2 \rho ^2 v^2
- S^{valley} (z)),
\label{modello}\eeq
 by the
saddle point method (note that they commuted the Wick rotation with $Im$).
The saddle point equation  relates the
relevant values of $\rho $ and $R$ to the initial energy $E$.

They find that the Holy Grail function  increases monotonically  and
reaches precisely zero value (hence no exponential suppression of
baryon number violation) at an energy of order of the sphaleron energy!
($x=x_{KR}\equiv 8\sqrt{3/5}$). See Fig. 4.

This result is repeatedly  referred to  in the literature as a very
encouraging sign that the $\Bvio$ cross section might reach the
geometrical size at high energies:
after all, if a {\it reasonable} dynamical model gives
 an interesting result,  why could  it not be true also  in  the
  real world?
    Unfortunately,  there  are strong reasons to suspect
  that the assumption {\it (ii)}
 used by Khoze and Ringwald is too naive and
  is hardly justified, precisely for values of $R/\rho  \le 1$  where
  the valley action sharply drops to zero.
We shall come back to the discussion of this model towards the
end of  Section 5.

Note that, by construction,
 the valley field depends on the choice
of the weight function $w$: other choices might possibly
introduce   different kind of  difficulties such as the
"bifurcation"
(\ie the loss of the saddle point of
 Eq.(\ref{modello}))\cite{dor92}\co\cite{suryak91}\co\cite{papero}.
A Minkowskian space formulation of the valley method was also
given\cite{bal91}.

Finally,  let us  mention the  recent perturbative
calculations of the Holy Grail function.\cite{Diakonov91}\daa\cite{silvestrov}
Balitsky and Sch\"afer\cite{bal92}  use the valley method (they check the
result by  using the effective Lagrangian approach) to compute the
third term of the Holy Grail
function, with the result:
\beq
F(x)=-1 +{9 \over 8}({E\over E_0})^{4/3} +{3 \over 16}
({E\over E_0}) ^2 + {3 \over 32}(4-3{M_H ^2 \over M_W ^2})
({E\over E_0})^{8/3}\log({E_0 \over E}).
 \label{holypert}\eeq
where  $E_0\equiv \sqrt{6}\sp$  so that $E/E_0 = x/\sqrt{6}.$
Silvestrov instead  uses  the $R$ - term method   to recover the coefficient
of
$x^{8/3}\log(1/x)$ , finding however only the piece (which agrees with
Balitsky and Sch\"afer\cite{Diakonov91}) depending on the
 Higgs and W boson masses. Diakonov and Polyakov\cite{Diakonov91}
 get the other piece, but with a factor $1/2$ compared to
Balitsky and Sch\"afer (these authors however work in the pure gauge sector).


\section{In search of  the Holy Grail.}

In this section several recent attempts to
compute  nonperturbatively the full Holy Grail function will be
reviewed in some details.

\subsection{Initial corrections and attempts  for a modified semi-classical
approximation}

Let us start with the discussion of  corrections involving high-energy
initial particles, \ie,  corrections to Eq.(\ref{growth})
due to  interactions between  hard particles or between hard and soft ones,
(this definition can be applied indifferently both
in the valley  and   $R$-term approaches). The lowest  hard-hard quantum
correction in the simplified case of bosonic particles
is shown in Fig. 5, and  hard-soft (hard-hard) corrections in the fermionic
case   in Fig. 1c (Fig. 1d).

First let us make some clarification on the terminology frequently used
in the literature (hence adopted here too).
The contribution to the amplitude
of each type of correction ("hard-hard",
"hard-soft" and "soft-soft") is not  uniquely defined
{\it by itself}\,; only the sum is well defined.
The ambiguity essentially arises in both approaches from the
 arbitrariness of the
choice of the scalar product (\ie \, the weight) with respect to which quantum
fluctations $A^q$ are taken to be  orthogonal to the streamline
$\de _\alpha A^{valley}$, see Eq.(\ref{valeq})
(valley approach) or to the zero modes $z_i$
of the second variation of the action
around the instanton (in the $R$-term approach).  This problem is often
referred to simply as the "constraint dependence" of each type of contribution.

In  fact,  from the usual insertion
of unity in the functional integral,
\beq
1=J_{w} \int d \xi \,\delta ( \int d^4x\, w(x) A^q (x) f(x) ),
\eeq
 (where $f=\de _\alpha A^{valley}$  or $f=z_i$ depending on  the
approach),
 a fictitious dependence on the weight is introduced. Such a dependence
is expected to  disappear
 order by order in perturbation theory only if all types of
 contributions  are added together\cite{mat91b}\co\cite{kleb91b}.

Single type of contributions thus in general depends on
 the choice of weight.
Mueller\cite{mu91}  however  argued that  soft-soft terms
are constraint-independent up to  order $x^{2}$.
On the other hand,
Khlebnikov and Tinyakov\cite{kleb91b} proved (in the $R$-term approach)
  that, already at the
$ x^{8/3}$ order, the soft-soft contributions to the Holy Grail function,
$F^{ss}(x)$, are ambiguous: they showed that an appropriate variation of the
weight induces a variation of the corresponding Levine-Yaffe\cite{ly}
propagator of quantum fluctuations giving a modification of $F^{ss}$ to
this order. A similar conclusion were obtained in the valley approach
by Arnold and Mattis\cite{mat91b}.
Out of  this observation comes also  a suggestive idea\cite{kleb91b}  that
if soft-soft corrections exponentiate but are ambiguous, perhaps
 the initial
state corrections must  also exponentiate to eliminate the ambiguity.
As regards  $O(x^{8/3})$ soft-soft term,
the constraint dependence would be removed by hard-soft corrections only,
since  hard-hard ones are known to  contribute starting  from $O(x^{10/3})$
in the low energy expansion
 (see below).

Let us now concentrate on the study of initial-state corrections,
 considering the  case of bosonic external particles
in the one instanton sector, as often done in
the literature.
The leading order amplitude for a $2\rightarrow n$ bosonic process
can be written as
\beq
A_{2\rightarrow n}= \int d\mu \, {\cal R}(p_1) {\cal R}(p_2)  \prod _{j=1,n}
{\cal R}(k_j) \,e^{-S_c}
\eeq
where $d \mu  $ stands for integrations over collective coordinates,
\beq
{\cal R}(p)\equiv\lim_{p^2\rightarrow m^2}(p^2-m^2) A^{inst}(p)
\eeq
are the on-shell residues of the (Minkowski analytically
continued) instanton field and $p_1,p_2$
are the hard momenta.

The first hard-hard correction
 comes from connecting together two hard particles with
a propagator (while other soft lines are kept unchanged), see Fig. 5.
  It
yields a correction
\beq
\delta A_{2\rightarrow n}= \int d\mu  D(p_1,p_2)  \prod _{j=1,n}
{\cal R}(k_j)\, e^{-S_c}
\label{dA}\eeq
where
\beq
D(p,q)\equiv\lim_{p^2,q^2\rightarrow m^2}(p^2-m^2)(q^2-m^2)\, G(p,q).
\eeq
$G$ is the Fourier transformed (Minkowski-continued) propagator
of the boson fields in the instanton background (\ie the constrained
inverse of the
second variation of the action evaluated at the instanton field,
which we call $\Box_c $). It
satisfies  the Levine-Yaffe equation\cite{ly}:
\beq
\Box_c G(x,y) = \delta (x-y)  - \sum_{i,j} f_i \,(\Om^{-1})_{ij}\, z_j
\eeq
arising from constraining the quantum fluctation to be orthogonal to
general constraint $f_i$ (having non singular overlap matrix
$\Om_{ij}= \int dx \,f_j(x) \,z_i(x) $ with the zero modes of $\Box_c$,
$z_j$). Clearly, for  $f_i=z_i$ one  recovers the usual BCCL
propagator\cite{bccl}.

We are thus interested in the behavior of the  double residue of
the  propagator, $D(p,q)$, in the kinematical
limit of interest  $pq\rightarrow \infi$, and $p^2=q^2=m^2$.
Explicit calculations in the O(3) two dimensional sigma
model\cite{krt91c}, and in pure $SU(2)$ gauge theory \cite{mu90} yields
\beq\label{factorization}
D(p,q)\rightarrow -c g^2 \rho^2 (pq)\log (-pq) {\cal R}(p) {\cal R}(q)
 + O( g^2 (pq)^0)
\eeq
 where $R$ are the residues of the instanton fields, and c is a
positive (model dependent) constant.
Applying the Regge-pole
tecnique to the operator $\Box_c$,  Voloshin\cite{vol91c}
 showed
that the constant c is
related to the translational  zero modes of $\Box_c$, $z_\mu
\propto \de_\mu A^{ist}$:
\beq
c= (z_\mu , |x|^2 z_\mu )^{-1}.
\eeq
 Due to the fast growth of  $D(p,q)$ with the scalar
product $(pq)$  (Eq.(\ref{factorization})),  this
enhancement overcomes the suppression factor $g^2$
when $(pq) \sim (M_W/g^2)$.  The correction  Eq.(\ref{dA})
is no longer suppressed compared to the leading term (an  estimate, $\rho \sim
M_W$ is used).

In a subsequent work\cite{mu91}  Mueller furthermore showed
 by direct evaluation that, for a pure gauge
theory,
 the quantum correction to the two (hard) point Green function
up to the order  $\alpha^N$ are of the form
\beq
\sum_{r=1}^{N} {1\over r !} [( -cg^2\rho^2  (pq)\log (-pq))^r R(p) R(q)
+ O(g^{2r}(pq)^{r-1}) ]
\label{hardcor}\eeq
where the leading terms of order $g^{2r}(pq)^{r}$
 come from the "squared tree" diagrams of Figure 6.
The dominance of the squared trees graph
is justified by rapidity ordering arguments\cite{mu91}$^,$\cite{mattis}.
 For an alternative (simpler) derivation of this result
 see the work of Li et al.\cite{mcler91}.

If one takes only the leading terms of Eq.(\ref{hardcor}) and sums the
series up to $N =\infi $
 the hard-hard terms exponentiate into a
factor
\beq
e^{-cg^2\rho^2 (pq) \log (-pq)}\label{naive}\eeq
which gives a contribution of order $O(\e ^{10/3})$ to the Holy Grail
function\fnm{d}\fnt{d}{Note that the condition of factorization
$E/N \gg 1/\rho $ (high energies in each final  branch of the tree graphs)
 and of exponentiation
$N\gg g^2\rho^2 (pq)$ seems to force $\e \ll 1$  }.
This (naive) exponentiation
could be taken as a first concrete  hint suggesting
the possibility of taking into account
   the initial
corrections semi-classically.

\smallskip
The first attempt for  a semi-classical treatment of initial
corrections was done by Mattis, McLerran and Yaffe\cite{mcler92}.
Their idea consists of including the initial  particles into
 an effective action $S_{eff}$ by means of the  identity:
\beq
\int dA \,A(p_1) A(p_2) \prod_{j=1,n} A(k_j)\, e^{iS(A)}
=\int dA \prod_{j=1,n} A(k_j)\, e^{iS_{eff}(A)}
\eeq
where
\beq
S_{eff}(A)\equiv S(A)-i \sum_{i=1}^{2} \log A(p_i),
\eeq
(the soft fields $A(k)$ can be  be treated with the usual $R$-term method).
The solution of the (non covariant)
field equation coming from the variation of $S_{eff}$,
\beq
{ \delta S \over \delta A(x)} =i  \sum_i e^{ip_ix}/A(p_i),
\eeq
could  in principle be used to give  a semi-classical
approximation that includes the leading
initial effect for $\e \sim 1$
\fnm{e}\fnt{e}{Also in this case, if the Minkowski propagator had
a singular behavior, the loop contributions could be important. Approximating
the  solution by the usual instanton and doing
perturbations around this (wrong) background, the authors found that
 a cancellation arises
between potentially large one-loop terms and  that tree contributions reproduce
the Mueller's corrections: this suggests that  in the correct
background loops are negligible.} .

\smallskip
Another suggestion  was made by  Mueller\cite{mu92}.  The   main
idea is  to exploit the arbitrariness in the choice of the constraints for
quantum fluctations, (in the LY propagator) to simplify the form of initial
state interactions as much  as possible.
He derived the behavior of the generic LY propagator in the kinematical
(Minkowskian) region of interest ( \ie  $(pq)\rightarrow \infi $ and on shell)
and was able to find  an appropriate  (noncovariant) choice of the constraints
that eliminates the leading term of $O(pq \log (-pq))$
 of the LY propagator.  This choice
consequently  eliminates all the leading multi-loop hard-hard corrections.
As for  the hard-soft interactions, Mueller demonstrated, by using  a
diagrammatic analysis that, for a single soft particle the  interactions
with hard particles can be included
 just by substituting  the instantonic background
by  the solution
 $W_\nu^a$ of the (Minkowskian) Y.M. equation  with a source term:
\beq
D_\mu^{ab}(W)\, G_{\mu\nu}^b(W)=J^a_{\nu}(W)-{1\over \xi}D_\nu^{ac}(A)\,
D_\mu^{cb}(A)\, W^b_{\mu}
\eeq
where $D(W),D(A)$ are covariant derivatives respectively in $W$ or in  the
instanton
background, $ G_{\mu\nu}^b(W)$ is the field strenght in terms of $W$
and $\xi $ is the usual gauge parameter.
 The source $J$ contains in a  complicated (nonlocal) way the
space-time integral of the solution $W$, and has a singularity structure that
makes  the continuation  to Euclidean time  impossible.
For more than one soft particles, hard effects are to be taken into
 account with the
usual $R$-term scheme (all tree graphs should be summed) in the new background
$W$.
A Euclidean valley resummation does not seem to be  applicable due to the
intrinsically Minkowski singularities.

\smallskip
It is not clear at present how useful
 the two approaches mentioned here might be
 in practice, as  they (both) involve the solution
of an extremely difficult (non-covariant) Yang-Mills equation with a
 source depending in a complicated manner on the solution itself
\fnm{f}\fnt{f}{One must also check, in the context of baryon number
 violation, that the solution has nontrivial topological number.};
 they are furthermore
incomplete in the sense that they must be complemented with some method for
 resumming soft trees.
 On the other hand,  these innovating  works  perhaps tell us
 that a semi-classical approach including  initial corrections is in
 principle possible.

\subsection{Multiparticle approach}

A more promising approach for  estimating the complete
 Holy Grail function (i.e. including initial state corrections)
is that of  Rubakov and Tinyakov\cite{tin92}$^,$\cite{tin92b}.
The  idea is to compute first the inclusive cross section
$n_i\rightarrow \all$  for large  $n_i$  ($n_i={\nu_i \over g^2}$), for
which  a semi-classical method is likely to be applicable
straightforwardly (see below). Then $ \nu_i $ can   be  sent to zero
{\it afterward} in
$F(\e ,\nu_i)={g^2\over 16 \pi^2}
\log \sigma_{n_i\rightarrow \all},$   assuming that  the limit is smooth,
 $g^2\log\s2 \sim \lim_{\nu_i \to 0} F(\e ,\nu_i)$.

 Note that while for $\s2$ the exponentiation of quantum (loop) corrections
 would appear rather miraculous, it is quite reasonable that
 multiparticle inclusive cross section can be evaluated semi-classicaly
 (this will be proved below).

 There are two ingredients for such semi-classical calculations.  First,
 an appropriate Minkowski  boundary conditions (taking account of the quantum
 number, energy, etc. of the initial state) must be used.  This is to be
compared  to the  standard calculation  of perturbative  S-matrix
 elements where the vacuum (Feynman) boundary condition is used.  For
 this purpose the coherent state  representation of the S-matrix turns
 out to be particularly suited.

 Secondly,  the fact that one is interested in  nonperturbative,
 classically forbidden processes, forces one to consider the time evolution
 partially in Euclidean direction (see below).   As a result the
 classical equation (and its solution) will be defined  along a complex
 time contour.

For completeness, let us first take a few steps back and review briefly the
 coherent state representation for the S matrix (Khlebnikov et
 al.\cite{krt} have  used this method extensively  in formulating  the
 $R$-term method.)

First consider  the one-dimensional  harmonic oscillator,
(generalisation to many-degrees of freedom is straightforward).
More details can  be found in textbooks\cite{coherent}.
In terms of the vacuum state $|0\ket $  and creation operator
$A^{\dagger}$, a (non-normalized) coherent state can
be defined, for every complex number $a$ as
\beq |a\ket =e^{A^{\dagger} a}\,|0\ket . \eeq
It is   easy to derive from commutation relation $[A,A^{\dagger}]=1$
the following  properties:
\bea
A \,|a\ket &=& a |a\ket \label{coerdef}\\
\bra b | a \ket & =& e^{b^* a}\label{coerscal}\\
e^{f A^{\dagger} A} |a\ket &=&|a \,e^{f} \ket\label{coerprop}. \eea
The wave  function of a coherent state in position representation is
also useful:
\beq \label{wavef}
\bra  q| a\ket = e^{- \half a^2- \half \om q^2 + \sqrt{ 2 \om}a q},\eeq
(it can be checked that this satisfies Eq.(\ref{coerdef})
when $A$ is expressed
in  the coordinate representation: $A={1\over \sqrt{2}} (\sqrt{\omega}q
+{1\over\sqrt{\omega}}{\de \over \de q})$).
The  action of an operator $S$ in the coherent state representation is
obviously given by its matrix elements: defining the kernel $\S (\bb, a) \equiv
\bra b  |S| a \ket ,$
\beq \bra b| S|\psi\ket =\int {da\, d\ba \over 2\pi i} e^{-\ba a}
\,S (\bb, a) \,\psi (\ba ),\eeq
where the exponential takes into account
the non-normalization of our  coherent states. Note also that in the integral
we must fix $\bar{a}=a^*$.

Finally note the following relation involving the matrix element of an operator
$S$ between fixed number of quanta and its kernel:
\beq\label{element}
\bra n |S| m\ket = \de_{a}^m  \de_{\bb}^n \S (\bb,a) |_{a=\bb=0}.\eeq

That is  all one  needs to construct the coherent state representation of the
 scattering matrix (in the Fock's space of asymptotic states)
\cite{coherent}.
 Starting from the usual interaction representation of $S$
\beq S=\lim_{T_i \rightarrow -\infi}\lim_{T_f \rightarrow \infi }
e^{i H_0 T_f} U(T_f,T_i) e^{-i H_0 T_i},
\qquad (H_0=\int d\k \,\omega _{\bf k} A^{\dagger} _{{\bf k}} A_{{\bf k}} )
\eeq
and introducing the  resolution of unity in terms of fields at $T_i$, $T_f$
(which allows to use the path integral representation for
$\bra \phi_f |U(T_f,T_i)|\phi_i\ket $), we obtain the following expression
for the kernel of the $S$ matrix:
\beq\label{kernel}
S (\bb,a )=\int d\phi_i \, d\phi_f \, d\phi \,
e^{B_i(\ak,\phi_i)+B_f(\bbk ,\phi_f)} e^{iS(\phi)}
\eeq
where the boundary terms
\bea
B_i(\ak ,\phi_i)&\equiv&-\half \int d \k \,\ak \amk e^{-2i\omk T_i}
	-\half 	\int d \k \,\omk \phi_i(\k )\, \phi_i(-\k ) \non\\
	& &+\int d\k \sqrt{2\omk}\,\phi_i(-\k ) \,\ak\, e^{-i\omk T_i}
	\label{bi}\eea
\bea
B_f(\bbk ,\phi_f)&\equiv&-\half \int d \k \,\bbk \bbmk \,e^{2i\omk T_f}
	-\half 	\int d \k \,\omk \,\phi_f(\k )\, \phi_f(-\k ) \non\\
	& &+\int d \k \sqrt{2\omk}\,\bbk \,\phi_f(\k ) \,e^{i\omk T_f}
	\label{bf}
\eea
come from wave functions of coherent states (see Eq.(\ref{wavef}), and also
Eq.(\ref{coerprop})).

\smallskip

In  the multiparticle
 approach of Rubakov and Tinyakov\cite{tin92},
one    actually considers another, slightly different kernel: the
 idea is to project
the initial coherent state  onto the eigenstate of  an
operator
\beq\label{operator}
\O =\int d\k f_{\k }A^{\dagger}_{\k } A_{\k } \eeq
(typically the energy) with  a fixed eigenvalue $\Om$
(generalization to cases
with  more
than one operator is straightforward).
One is then interested in:
\beq
\S_\Om ( \bb ,a )\equiv \bra b| S \P_\Om |a\ket .\label{kernel1}\eeq
Substitution  in Eq.(\ref{kernel1}) of the integral representation
of the projector
\beq
\P_\Om = \int d \xi\, e^{i(\O-\Om )\xi}
\eeq
and use of property Eq.(\ref{coerprop}),  yield
\beq \label{esseomega}
\S_\Om ( \bb ,a )=\int d \xi
\int d\phi_i \, d\phi_f \, d\phi \,e^{-i\Om \xi}\,
e^{B_i(\ak e^{i f_{\k} \xi}  ,\phi_i)+B_f( \bbk   ,\phi_f)}\, e^{iS(\phi)}
\eeq
where the function $f_{\k}$ refers to the operator $O$, Eq.(\ref{operator}).

A quantity of interest, then,
is the transition rate for
the microcanonical ensamble (respect to $\O $). It is obtained by
summing  the inclusive cross sections  over  the initial  states
 with a
 definite value of $\O $,
\beq \label{sigmaomega}
\si  (\Om )=\sum_{i,f}|\bra f|\S \P_{\Om }|i \ket |^2. \eeq

(Actually, there is a slight
lack of precision in this expression.  One is really
 interested here in a partial sum over  B- (and L-) violating final states
$f$, so that the unit operator part
 of the S-matrix actually drops out. Otherwise,
Eq.(\ref{sigmaomega}) would have no physics contents!
Formally, however,  the sum over $f$
is done without
 any restriction (see the next equation).   The restriction over selected
final states is to be taken into account
 by a judicious choice of the classical backgrounds.)

By substituting  Eq.(\ref{esseomega}) in Eq.(\ref{sigmaomega}), one obtains:
\beq \label{sigmaomega2}
\si (\Om )=
\int d \phi \,d \phi '\, d a \, d \ba \, d b\, d \bb \,d \xi\,
e^{W}
\eeq
where
\bea
W&=&-i\Om \xi - \int d\k\, \bbk \bk - \int d\k\, e^{-if_{\k } \xi}\,\bak \ak
+ B_i(\ak ,\phi_i)+B_f(\bbk ,\phi_f)+iS(\phi )\non\\
& &+ B_i(\ak ,\phi_i')^*+B_f(\bbk ,\phi_f')^*-iS(\phi ')^*. \label{W1}\eea
Note that a  rescaling $\ak \rightarrow \ak e^{-if_{\k }\xi}$,
 $\bak \rightarrow \bak e^{if_{\k }\xi'}$,
 the integration over $d(\xi +\xi')$ as well as the substitution
$\xi-\xi'\to\xi$ have been made.

The nice feature of   $\si (\Om )$ is that the functional integral
appearing in Eq.(\ref{sigmaomega2}), although very complicated,
can be evaluated  semi-classically
 in the limit of small $g$, provided that the external parameters $\Om$
are of the order of ${1 \over g^2}$. Indeed, rescaling as
\beq
(\phi,\phi',\ba,a,\bb,b)\rightarrow
(\phi,\phi',\ba,a,\bb,b)/g
\eeq
one finds  $W={{\tilde W}\over g^2}$, where ${\tilde W}$ depends on the
rescaled fields, on $g^2\Om$
and on $M_W$,
but not explicitly on $g$.  Thus the usual saddle point tecnique can be
applied, yielding  a result of the form
\beq
\si (\Om ) =\exp \{ {1\over g^2} F(g^2\Om,M_W)\}.\eeq
The saddle point field  will be a solution of a classical
equation of motion, with  some new type of boundary conditions
(different from those in the usual vacuum to vacuum transitions)
 involving the saddle
point values of the  complex fields  $a,b$; more on this later.

(The present  formalism was actually introduced earlier by
 Khlebnikov, Tinyakov, Rubakov\cite{krt91b}, in which the operators
$O$ were chosen to be the four momentum. They considered the microcanonical
ensemble of initial states with a fixed momentum $(E,\vec{0})$, with the
idea that  the new boundary conditions  will be
more physically relevant in processes of production of many particles,
as they keep memory of the incoming and outcoming fields.
The relevant saddle point configuration is  an analytic
continuation to an adequate complex time path of a real Euclidean
time configuration,   the so-called  "periodic instanton".  $\si (E)$
however turns out not to be directly related to the cross section
 $\sigma _2$  one is interested in\cite{krt91b}.)

 The main new idea of Rubakov and Tinyakov\cite{tin92} is  to consider
the  cross section for the initial
 microcanonical ensemble
with fixed  energy $E$ {\it and}  fixed number of incoming
 particles $n_i = \nu_i/g^2$,
$\si (E,n_i)$.

The advantage  of this  approach is that for $\nu_i \neq 0$ all
 the initial states
corrections are automatically taken  into account in the semi-classical
 approximation,
since the initial particles  are represented as fields appearing in the
boundary terms.
Thus, if  $\s2$  is recovered in
 the limit $\nu_i\rightarrow 0$, it is likely that
 the initial corrections are accounted for: this would reduce the
problem of evaluating the Holy Grail function effectively to the search
of an adequate
classical configuration.

 It was also shown\cite{tin92b} that, perturbing the functional
integral for $\si (E,n_i)$ around the instanton (expansion valid for
$E\ll\sp$),
that
 the leading hard-hard corrections thus inferred do coincide with the naively
exponentiated Mueller's series, Eq.(\ref{naive}).

\smallskip

Let's us consider thus  the expression Eq.(\ref{sigmaomega2}) for
$\si (E,n_i)$ (with the obvious generalizations: $d\xi\rightarrow d\xi
d\eta$, $\Om \rightarrow E,n_i $, $f_{\k }\rightarrow \omk ,1$), and discuss
the boundary conditions\cite{tin92c} that characterize the saddle point
field.

The integration of the final states $\bb, b$ gives exactly a
term $\delta (\phi_f-\phi_f')$ (by completeness of  coherent states); then the
subsequent integration over $\phi_f,\phi_f'$
(keeping into account of all boundary
 terms\fnm{g}\fnt{g}{Note that  also a term
$${i\over2}\int d\k\,
(\phi_{{\bf k}} {\dot \phi}_{{\bf -k}} -\phi_{{\bf k}} '
{\dot \phi'}_{{\bf -k}})|_{T_i}^{T_f}$$
 in the exponential coming from integration by parts of the kinetic term
of the fields' action, are essential to the integration
over initial and final fields. }), gives a
$ \delta (\dfi_f-\dfi_f')$. Thus, being $\phi=\phi '$,
$\dfi={\dot\phi}'$ at $T_f=+\infi $ the two classical solutions coincide
over all Minkowskian space.

The integration  over $\ba,a$ is Gaussian and can be done exactly; the
 integration  over $\phi_i,\phi_i'$  leads to  the following boundary
 conditions at $T_i\rightarrow -\infi$ (the
saddle point values for $\ba,a$ have been substituted already):
\bea
i\,\dfi_i(\k )+\omk \phi_i(\k )&=&e^{i\Delta_{\k }}
 	[i\,\dfi_i'(\k )+\omk \phi_i'(\k )]\label{bc1}\\
i\,\dfi_i(\k )-\omk \phi_i(\k )&=&e^{-i\Delta_{\k }}
 	[i\,\dfi_i'(\k )-\omk \phi_i'(\k )]\label{bc2}
\eea
where $\Delta_{\k }\equiv \omk \xi + \eta$.
Subsequent saddle point evaluation of  the integral over  $\xi $ and $\eta$,
 will then fix
the parameters of the solution in terms of $E,n_i$.

By time traslation invariance, the variable
 $\xi$ associated with energy can always be taken as purely imaginary:
$\xi=i\xi_0$; perturbative calculations\cite{tin92b}$^,$\cite{tin92c}
suggest that also the saddle point value of $\eta$ is purely imaginary; so
we can choose $\eta=i\eta_0$.
 It was furthermore observed\cite{krt91b}$^,$\cite{tin92c}
that the parameter $\xi_0$ can be removed from boundary conditions
Eqs.(\ref{bc1}-\ref{bc2}) by choosing suitable contours in complex time
plane: one choice could be the one shown in Fig. 7 (with $\xi_0=BB'$).

If we assume the solution to be  free at
$\t=\re \, t\,\rightarrow -\infi $,
and  write in this limit
\bea
\phi (\k )
&=&f_{\k }\,e^{-i\omk \t }+\bar{f}_{-\k }\,e^{i\omk \t }\qquad t \in {\rm AB}
	\label{free1}\\
\phi ' (\k )
&=&g_{\k }\,e^{-i\omk \t }+\bar{g}_{-\k }\,e^{i\omk \t }\qquad t \in
{\rm A}^{\prime} {\rm B}^{\prime},
	\label{free2}
\eea
substituting in   Eqs.(\ref{bc1}-\ref{bc2}) we obtain  simplified
boundary conditions:
\beq \label{bc3}
 f_{\k }= e^{-\eta_0}g_{\k };\qquad \bar{f}_{\k }= e^{\eta_0}\bar{g}_{\k }
\eeq
which relates asymptotic positive and negative frequency components on the
two branches of the path, $\hbox{AB}$ and $\hbox{A$'$B$'$}$.

As anticipated before, the necessity of using the complex-time deformed
path comes out from the double requirement of having
free particle incoming (outcoming) at Minkowskian time $\re t \rightarrow -
(+) \infi $ and {\it of describing  a semi-classical, classically
 forbidden transition}, that requires a part of the evolution to proceed
in the Euclidean  time.

The idea of the complex-time deformed paths is not really new: it is known
 (and   was invented first) in   quantum
mechanics, in the W.K.B. approximation for  the fixed energy Green's functions
in the case of barrier penetration\cite{mclau}.  In that case, the relevant
classical solution
 is composed of  Minkowski solutions for the initial and final parts of
 evolution  and of an
intermediate,  Euclidean  solution describing the tunnelling.

Let us  now have a  closer look at the required properties of
the classical field we are looking for.
If the solution to field equations  satisfying boundary
conditions Eq.(\ref{bc3})
 is unique, then it must obey the relation $ \phi(t^*)=\phi^*(t) $
(coming from reality of field equations);
that implies the reality of the field on the positive Minkowski time. It
follows from
 Eqs.(\ref{free1}-\ref{free2}) that
\beq
f_{\k }={\bar g}_{\k }^*;\qquad  g_{\k }={\bar f}_{\k }^* .
\eeq
Also,  it can be seen that the saddle point
values of $a,\ba $ are complex conjugate  of each other, which
 means that the sum
over  the initial states is dominated by a single coherent state.

Perturbative calculations\cite{krt91b,tin92c} support  the    conjecture
that on the Euclidean time piece the solution might be real: this (together
with the reality on the positive real time axis) would imply the presence of
a turning point $\dfi =0$ for $t=0$\cite{tin92c}.

All these hypotheses,  put together, imply that  the required saddle point
configuration,
 determining  the microcanonical cross section $\si (E,n_i)$ would be
a solution of classical equations on the complex time path of Fig. 7
with boundary conditions:
\bea
\dfi |_{t=0}&=&0;\non\\
\phi (\k )&=&f_{\k }\,e^{-i\omk \t}+e^{\eta_0}f_{-\k }^*\,e^{i\omk \t}
\label{bc4},\eea
where $ \t =\re \; t\rightarrow -\infi $.

\smallskip
Unfortunately,  the "microcanonical" saddle point is not known at present
even numerically  at energies of order $\sp$.
A perturbative expression for
this solution (valid at low energies) is known \cite{tin92c} in the context
of two dimensional Abelian Higgs model: it looks like a chain of alternated
 instantons and  anti-instantons.
This study put into evidence the problem
of the presence of singularities of analytically continued
classical solutions in the  complex time, that
 the chosen
path must not  touch. See the discussion below, Eq.(\ref{SQN}).

\smallskip
A slight variation of the approach of Rubakov and Tinyakov,  is
to compute the  cross section for a {\it given} coherent state of fixed energy
\beq \label{sigmaea}
\si (E, \{ \ak \})=
\int d \phi \,d \phi ' \, d b \,d \bb \,d \xi \,d \xi'\,
e^{W}
\eeq
with
\bea
W&=&-iE (\xi -\xi ') - \int d\k \,\bbk \bk
+ B_i(\ak e^{i\omk \xi},\phi_i)+B_f(\bbk ,\phi_f)+iS(\phi )\non\\
& &
+ B_i(\ak e^{i\omk \xi '},\phi_i')^*
+B_f(\bbk ,\phi_f')^* -iS(\phi ')^*, \label{W2}\eea
where now $a, \ba$ are the arbitrary but fixed complex numbers.
 Integration over final fields goes as before.  What changes
  is the integration over  $\phi_i,\phi_i'$ that
by saddle point evaluation give the initial boundary conditions:
\bea
i\,\dfi_i(\k )+\omk \phi_i(\k ) &=& \sqrt{2\omk}\, \ak \,
 e^{-i\omk (T_i-\xi)}\non\\
i\,\dfi_i'(\k )-\omk \phi_i'(\k ) &=& -\sqrt{2\omk} \,\ba_{-\k }
\,
e^{i\omk (T_i-\xi')}. \label{bc5}
\eea
Clearly, if one would have known the solution of the boundary
 condition problem
for {\it generic} values of $a,\ba$  one  could estimate semi-classically
 Eq.(\ref{sigmaea}) and
recover directly the two particle
cross section by differentiating twice  $\si (E, \{ \ak \})$ (see
Eq.(\ref{element})).
\smallskip

Unfortunately, the resolution of the general boundary condition  problem
is a  hard task. Some recent
 works\cite{tin93}$^,$\cite{gould93}$^,$\cite{kya92}
pursue  a more modest goal of calculating semi-classically
$\si (E, \{ \ak \})$ with $a,\ba$ fixed {\it a posteriori} to
satisfy the
boundary conditions Eq.(\ref{bc5}), starting from   some  given
 classical solution.
These works have many common features.
 They study exact solutions  of
field equations in conformally
 invariant models\fnm{h}\fnt{h}{They are all massless models that are
supposed to describe
the high-energy dynamics of the massive ones. Clearly with this approach it
is impossible to recover the low-energy behavior of massive models.},
 obtained by some symmetry
requirement that simplifies  the problem to an equivalent one dimensional
one.
   In particular, the first\cite{tin93}
 studies the massless two dimensional $O (3)$ sigma model,
  searching for $O (2)$
symmetric solutions; the second\cite{gould93} considers the
$SO(4)$ conformally invariant solutions of Minkowskian $SU(2)$
pure Yang Mills theory,
the so-called L\"uscher-Schechter solutions\cite{luscher}$^,$\cite{schecter};
the third\cite{kya92} studies an $O(4)$ invariant solution of a massless,
four dimensional $\phi^4$ theory with the positive ("right") sign of the
coupling constant.

All these solutions are  continued to a complex path as in Fig. 7,
and possess a turning point at $t=0$. It turns out that the
position of the path
relative to the singularities of the solution determines its very nature.
As  a significant example   note that Yang Mills complex time solution
is such that the Minkowskian  action $S$ and the topological
charge $Q$ evaluated
along the path, obey the relation
\beq
{g^2\over 8\pi^2}\im S= Q=N, \label{SQN}\eeq
 where $N$ is the  number of singularities between the path and
  Minkowski time
axis (relation that is identical to the usual  multi-instanton).
The path considered in his particular  case
 corresponds to the one instanton
sector.

Evaluating the average  number of initial (from the derived $a,\ba$) and
 final particles (from the saddle value of $\bb,b$), it turns out that these
solutions allow to compute  the probability of a process with parametrically
less particles in the  initial state than  the final one
(for example\cite{gould93} $\nu_f=\nu_i^{7/ 8}$ for small $\nu_i$).  But in
the limit $\nu_i\rightarrow 0$ the cross section is exponentially
 dumped by the full t'Hooft
suppression factor $\exp\{-{16\pi^2}/ g^2 \}$.
These solutions however do  not
maximize the transition probability in the one
 instanton sector at a given energy,
(because the resulting $a,\ba$ have not opposite phase, see above).
Thus these solution neither are directly related to the $2\rightarrow \all$
processes, nor can be used to establish a rigorous upper bound to that
cross section. Nevertheless they are a useful benchmark to understand the
property of more physically relevant solutions (such as the "microcanonical"
one considered before).

An important, nontrivial question, is if general  complex
gauge fields configurations
 defined on a complex contour, satisfy a generalized
Atiyah-Singer index theorem\cite{as}\co\cite{NIELS} of the form:
\beq
n_l-n_r = \int_{\rm path} \Tr F {\tilde F}=Q
\eeq
where $n_{l}$ ($n_{r}$) are the left (right) handed zero modes
of the Dirac operator.
Such a theorem would imply  that if we had a classical solution with
complex topological charge $Q=\pm 1$, we would be assured  the  presence of
the fermion zero mode which is  essential
for the nonvanishing of the anomalous cross section.

Also, one may ask if  there is  a level of the Dirac Hamiltonian crossing zero
in time evolution
 (physically interpretable as creation or annihilation of a  hole or a particle
depending on the direction, see Section 5), in one-to-one correspondence
 with the zero modes.
A step in the  direction of answering these questions,
 was taken  in a work by Rubakov and  Semikoz
\cite{rubakov}, where
the authors verified
  these conjectures  in the case of the
two dimensional Abelian Higgs model.

Let us  end  this section by a brief  mention to a recent  work of
Mueller\cite{mu92b}.
Instead of the  "microcanonical" cross section
$\si (E,\nu_i)$ of Rubakov and Tinyakov, the author considers
a slight variation, with the initial state
consisting of  the (equally weighted) sum of all pure coherent states
 with energy $E$
and with
$n_1=\nu_1/g^2$  right moving and $n_2=\nu_2/g^2$ left moving
 W's (he considers
only the pure gauge sector).
 The processes with such  modified initial
states  seem to be considerably closer kinematically to  those with
 the  two particle initial
states, than the simple microcanonical ones.
Subsequently the functional integral is expanded perturbatively  around the
single instanton.
An accurate study of the hard-hard corrections, (up to  $\e ^{10/3}$ in $F$),
and of hard-soft (up to  $\e ^{16/3}$) shows that in the limit $\nu_1=\nu_2
=\nu_i\rightarrow 0$  these contributions to
the exponent $F(\e ,\nu_1 , \nu_2 )$  join smoothly with the result for
 two particle initial state. This is quite nontrivial, and holds only
after  compensations of  divergent $1/\nu$ terms.

It would thus appear that   a semi-classical (perhaps
numerical) study of processes of the form $n_1+n_2\rightarrow \all$
provides  a good
way to estimate  nonperturbatively the Holy Grail function
for  the desired $2 \rightarrow \all$ transition.  How realistic this
possibility is in practice, is however  unknown.


\section{Quantum mechanical analogue problems}

In order to get an  intuitive,  physical   understanding of  our four
dimensional problem, several authors
studied  toy models in the context of one-dimensional quantum
 mechanics\cite{girard}\daa\cite{Diakonov93}.
The results of these studies will
  be reviewed briefly  in this section,
 following mainly the treatment of Diakonov and Petrov\cite{Diakonov93}.

Consider a quantum mechanical    double well coupled to a weak,
rapidly oscillating field.  The   Lagrangian is  given by
\beq
L={1 \over g^2} \left( {1\over 2}({dq \over dt})^2
-{1 \over 8}(q^2 -v^2 )^2
-Fq ( e^ {iEt} + e^{-iEt} ) \right) ,
\eeq
where $F$ is a small coupling constant.  One could  also introduce
 fermionic excitations (instead of the external
field), without however changing the results\cite{girard} essentially.

Let us ask now  the following question: what is the probability of the
transition
 from  the ground state in the left well to a highly  excited state in the
right one?
To  lowest order in the external field, the transition
probability from the ground state to   the $N$-th  state
is given by the well known Fermi's Golden Rule:
\beq
W=2\pi \delta (E+E_0 -E_N) |\q0n|^2 F^2
\eeq
where
\beq
\q0n\equiv\int dq \Psi ^{*}_{N} (q) q \Psi _{0}(q).
\label{matele}\eeq
The problem is then reduced to the calculation of the matrix element
of the position operator betwen two states with a large difference in
energy.

Two approaches have been taken to estimate  this matrix
element: one
 uses the semi-classical (W.K.B.) method
 (which will be  followed
below) and the other  (taken by Bachas) a more formal one.

 As is well known,
  the nonperturbed energy eigenstates of the double well
  potential are symmetric or antisymmetric under parity.
For a level lying much  below  the height  of the central
barrier, one can define
states localized in the left or right well:
\begin{eqnarray}
|N_L \ket & \equiv & {1 \over \sqrt{2}} (| N_s \ket + |N_a \ket) \\
|N_R \ket &  \equiv & {1 \over \sqrt{2}} (| N_s \ket - |N_a \ket)  .
\end{eqnarray}
It is clear however that when the energy is higher than the top of the barrier
these  states are no longer
localized in the left or right well.  In the Weinberg Salam theory in
four dimensions,  the states corresponding to different Chen-Simons numbers
are always
 well defined (for instance they can be distinguished by the number of
lefthanded
 fermions present).  Thus the analogy between the real problem and the
quantum mechanical toy model   is not self evident for
energies above the barrier.  But  the matrix element is exponentially
small  in such a  situation (see below)  so that the precise definition
of the final state is probably not crucial.

The matrix element is given, to exponential accuracy, by\cite{landau}
\beq
{\bra N_R |q|0_L\ket }=e^{(-S_1 +S_2 -S_3)},
\label{semicl}\eeq
where
\begin{eqnarray}
S_1&\equiv &
{1 \over 2 g^2 } \int_{-\infty }^{-v} dq(q^2-v^2 ), \\
S_2&\equiv &
{1 \over 2 g^2 } \int_{-\infty }^{q_D} dq \sqrt{(q^2-v^2 )^2 -\epsilon } \\
S_3&\equiv &
{1 \over 2 g^2 } \int_{q_E}^{q_F} dq \sqrt{(q^2-v^2 )^2 -\epsilon },.
\end{eqnarray}
where  $\epsilon \equiv 8Eg^2.$
The terms in the exponential are  shortened  action
with energy $0$ or $E$, corresponding to  the (complex time) trajectory
shown in Fig. 8.
To our purpose  the only important
contributions come from  the parts of the trajectory corresponding to
  Euclidean time evolution ($AB$, $CD$ and  $EF$ in Fig. 8),
 because the paths
in Minkowski time yield   only a phase factor.
Note  that when $E \rightarrow 0$ the contributions from paths
AB and CD tend to cancel each other while  the path EF gives
the usual tunnelling factor at zero energy.

As the energy increases,
it is clear that   $S_3$ diminishes
 but $S_1 -S_2$
grows with energy.  The problem is which  tendency
 dominates\cite{banks90}.

The  answer has been found for  very high and very low energies.
Furthermore, an argument has been given for the "half-suppression"
 i.e,  the   transition
 probability   at its maximum (which  occurs at  $E=v^4 / 8g^2$)
  being  the square root
of that at  $E=0.$\cite{kis}\co\cite{Diakonov93}

The transition probability   $W$ is proportional to the square of
the matrix element $\q0n$.
It  can be calculated without an explicit knowledge of the trajectory at
very high or very low energies.
Indeed, first consider the quantity
\beq
g^2 {d\log  W \over d\epsilon } =-g^2 {dS_1 \over d\epsilon}
+g^2 {dS_2 \over d\epsilon} +g^2 {dS_3 \over d\epsilon}.
\eeq
The right hand side can be simplified:
if $\epsilon < 1$ it is
\beq
g^2 {d\log  W \over d\epsilon} = {1 \over 2} {K(k) \over
\sqrt{1+\sqrt{\epsilon}}}
 \eeq
while  if $\epsilon >1$ it is equal to
\beq
g^2 {d\log  W \over d\epsilon}= {-K(l) \over 2\sqrt{2\sqrt{\epsilon}}},
\eeq
where $K(x)$ is the complete elliptic integral of the first kind\cite{tavole},
 $ k=\sqrt{{1-\sqrt{\epsilon} \over 1+\sqrt{\epsilon}}}$ and
$l=\sqrt{{\sqrt{\epsilon}-1 \over 2\sqrt{\epsilon}}}$.

After integration in $\epsilon$ one finds  the following results:
  \beq
g^2 \log  W = -{4 \over 3} + {\epsilon \over 8}(\log  {64 \over \epsilon} +1)
+O(\epsilon ^2)
\eeq
for $\epsilon << 1;$
 \beq
g^2 \log  W = -{\epsilon ^{3/4} \over 6}B({1 \over 4},{1 \over 2})
+O(\epsilon ^{1/4}),
\label{qmhg}\eeq
for $\epsilon >> 1$,
where $B(x,y)$ is the  Euler Beta function\cite{tavole}.

These results show that the transition probability  is  exponentially
suppressed
both at low energies (where W grows with energy) and at high energies
(where W is exponentially damped with energy).

To demonstrate the square-root suppression of the transition probability
 at its maximum, one  compares the
(imaginary) time along the paths CD and EF.
According to the classical mechanics
\beq
g^2 {dS_I \over d\epsilon }=T_I (\epsilon ),
\eeq
where $T_I(\epsilon)$ is  the time along the $I$-th path with energy $E$.
The time interval $T(\gamma )$ of the piece of the contour $\gamma $
  is thus given by
\beq
T(\gamma) = \int _{\gamma} {dq \over \sqrt{U(q)-\epsilon }}.
\label{interval}\eeq

Let us consider the difference of time intervals
 along the contours $\gamma _1$
and $\gamma_2$ (Fig. 9), in the complex q plane.
As long as the potential has no singularities for finite values of q, the
integrand of Eq.(\ref{interval}) has only
cuts at the classical turning points.
The cut being the square-root type, the value of the integrand at
both sides of the cuts differs only in   sign.
Using Cauchy's Theorem one gets then
\beq
T_1+T_2=2T_3
\eeq
or
\beq
-{dS_3 \over d\epsilon}=2{dS_1 \over d\epsilon }-2{dS_2 \over d\epsilon }
\eeq
and after integration in $\epsilon $, this yields
\beq
S_3 (\epsilon )=S_3 (0) -2S_1 (\epsilon ) +2S_2 (\epsilon ).
\eeq
At $\epsilon =1$, $ S_3$ vanishes,   thus
\beq
S_1 (1)-S_2 (1) = {1\over 2} S_3 (0).
\eeq
This result   connects the overlap integral (the left hand side of the
equation) to the tunnelling factor at zero energy, as anticipated.
  This is
 a very general result\cite{kis}\co\cite{Diakonov93},
 independent  of the
explicit form  of the double well used above.

Essentially the same result has been obtained by Bachas\cite{bac91},
 who  was able to give a {\it rigorous} proof of exponential
suppression of the induced high-frequency
transition amplitudes in an anharmonic potential
(one or two wells), based
on exact recursion relations between matrix elements of powers of the
position operator.
This demonstration settles the issue for the quantum mechanical problem,
but, in spite of optimistic expectations of the author, no generalization
to field theory is known, up to now.  Finally we find it  interesting that
the high energy
behaviour of the "Holy Grail function" (for
the case of the double well),
Eq.(\ref{qmhg}), found by
 the WKB method,  is consistent  with the exact bound found by Bachas, apart
from a logarithmic term which is probably beyond the precision  of the
semi-classical approximation.

 Landau's result, Eq.(\ref{semicl}),
 has    been earlier  generalized to field theory by
Iordanskii and Pitaevskii\cite{pit}.
The asymptotic behavior of the
imaginary part of the Fourier transform of a  two point Green
function  is given by:
\beq
Im \,G({\bf k},\omega) \sim e^{2( -\Delta S_I (0,0) + \Delta S_{II} (\omega ,
{\bf k}))}
\eeq
The two terms in the exponent correspond to the value of the action along the
trajectories at energy $E=0$ and $E=\omega$. This formula is valid
if $\omega$ and $k$ are large quantities, and
is similar to the quantum mechanical one.

The  method of Iordanskii and Pitaevskii  was recently applied
 to non Abelian  Gauge theories\cite{Diakonov93}.

By analogy to the quantum mechanical example one must   find
singular solutions to the classical Yang-Mills-Higgs equations of motion
at energy $0$ and $E$.
Diakonov and Petrov look for an $O(3)$ symmetric solution  in the pure
gauge theory.
At $0$ energy these are the instanton and  anti-instanton
with the size $\rho ^2$ changed to $-\rho ^2$, while it
is more difficult to find a solution if the energy is non
zero\cite{kleb91}.
In the limits of very high or very low energies an approximate
solution can however be obtained analytically.
In the low energy regime, the two contributions to the
reduced action corresponding to the path ABCD and EF in fig. 8
sum up and give the usual leading result Eq.(\ref{growth}).

In the high energy limit,  the cross section
for  isotropic multiparticle production  is found to
decrease exponentially with the energy as
\beq
\sigma (E) \sim e ^{- \const  ({\alpha \rho E
 }) ^{3/5}/\alpha }.
\eeq
(valid for fixed $\rho$ such that $\alpha \rho E \gg 1.$)
This result can be interpreted\cite{Diakonov93} in terms of an effective
strong
repulsive interaction between the "instanton" and  "anti-instanton" at
small separation.  Such a repulsion prevents the instanton pair from
collapsing, in contrast to what happens in the case of the valley (Section 5).
(See Klinkhamer\cite{klinkh} for a related discussion.)

The result should be correct also in the electroweak model, because both
at very high energies and low energies one can neglect the effects of Higgs
boson,
while at the sphaleron energy one should consider the complete
Yang-Mills-Higgs coupled equations.

A naive extrapolation of the results at intermediate energies shows
that  the maximum of the cross section is
achieved  near  the sphaleron energy where it is close to the square
root of its value at zero energy.
One cannot draw any conclusive answer because it is not clear whether the
particular classical solution used maximizes  the multiparticle
production cross section.

\smallskip
Finally  the work of Cornwall and Tiktopoulos\cite{tik92}\co\cite{tik93}
should be mentioned.  These authors use  a functional Schr\"odinger equation
approach to study  $SU(2)$ gauge theory,  exploiting  the analogy with
the quantum mechanical double-well problem.  First  it is argued that the
matrix element,
\beq \bra NE | \phi(0) | 0 \ket   = \int d\phi \, \psi^*_E\{\phi \} \phi
\psi_0 \{\phi \}    \eeq
(where the final state with energy $E$ consists of $N$ particles, $\phi$ is
the field operator), is bounded from above simply by
\beq  \psi_0 \{\phi_{NE}\}.   \eeq
$\phi_{NE}$ is the appropriate field configuration with quantum numbers
$E,$  $N.$
Secondly, the vacuum wave functional for the $SU(2)$ gauge theory
is known\cite{loos}  ($\phi = {\bf A}$):
\beq \psi_0 \{ {\bf A} \} = {\cal N} \sum_{J= -\infty}^{\infty} \exp\left[
-{2 \pi \over \alpha} |W({\bf A}) - J | \right],   \eeq
where $W({\bf A}) = {\cal N}_{CS}$ is the Chern Simons functional (see
Eq.(\ref{csn})).
By approximating the final state configuration ${\bf A}$ by the sphaleron
for which  $W({\bf A}) = 1/2,$  and by  keeping the $J=1$ term only,
these authors  get the half (or the square root)  suppression
 of $\Bvio$   cross sections.


\def\Ibar{\bar I}
\def\Cost{{96\pi^2\over g^2}}

\section {Unitarity  Bounds}

Various arguments based on unitarity, which all lead to the
"half-suppression" of the $\Bvio$ cross sections, are reviewed in
this section.

\subsection {Multi-instanton unitarization, "half-suppression",
"premature unitarization" and all that}
   The result of the original calculation of Ringwald\cite{ring90} and
 Espinosa\cite{esp90} violates unitarity if extrapolated
to high energies, for  reasons similar to those for which the Fermi
theory of weak interactions violated unitarity.  The
effective  interaction vertex
 is  given by a local, non-renormalizable
form\cite{svz}\co\cite{eff}\co\cite{bal92}:
  \bea   L_{eff} =
\int d x \int {d \rho\over \rho^5}\,\int d u  \,  d(\rho )
e^{-2\pi ^2\rho ^2{\bar \phi} (x) \phi (x)} \cdot \non \\
\cdot \left(
e^{ {2\pi^2i\over g} \rho^2 \Tr \{ \si_{\mu}\sb_{\nu} G_{\mu \nu}(x) \} }
+
e^{ {2\pi^2i\over g} \rho^2 \Tr \{ U\sb_{\mu}\si_{\nu}\bar{U} G_{\mu \nu}(x)
\}}
\right)
\label{Leff} \eea
where $U\equiv U_\mu \si_\mu$, $\bar{U}\equiv U_\mu \sb_\mu$ are the matrices
 representing a global orientation of the instanton in the isospin space,
and the instanton density $d(\rho )$
 contains $\exp (-S^{inst})$ and the functional determinant.

The use of Eq.(\ref{Leff}) in the lowest order
 to compute the S matrix elements will
necessarily lead to violation of unitarity.  Contributions
which are higher orders in Eq.(\ref{Leff})
- multi-instanton contributions -
have to be taken into account in order to restore
unitarity\cite{zak91b}\daa\cite{shif91b}.

Zakharov\cite{zak91b} (also  Aoyama and Kikuchi\cite{aoya90},
 Veneziano\cite{ven90})
 pointed out
 that such multi-instanton
 corrections can change
qualitatively the result found in the original  one instanton calculation.
A rough estimate of the multi-instanton contribution to the $2\to N$
amplitude, iterated in the
s-channel (see Fig. 10) and summed over the number of instantons,
  would give an answer,
\bea
A_{2 \to N} &\sim \sum_{k=0}^{\infty}
 A_{2 \to N}^{tree}\cdot (iA_{N \to N}^{tree})^{2k} \non \\
       &\sim A_{2 \to N}^{tree}/ \{1 + (A_{N \to N}^{tree})^2 \},
 \label{multi}\eea
where a  simple factorized form of $k$- instanton contribution is assumed,
the superscript "tree" indicates the single
instanton  (or anti-instanton) approximation,
and $N$ stands for the dominant,  multiparticle intermediate states
(it includes symbolically also the summation over them).  Although such a
simple, factorized structure of Eq.(\ref{multi}) is by no means obvious,
a more careful estimation of multi-instanton contribution will be
given in the next subsection  which will reproduce   essentially all the
results following from Eq.(\ref{multi}).

The single instanton amplitudes can be taken, for the purpose of
resummation, to be  of order, \beq
A_{2 \to N}^{tree} \sim  e^{-2\pi/ \alpha}\cdot K^{1/2},\label{tree2N}
\eeq
and
\beq
A_{N \to N}^{tree} \sim  e^{-2\pi/ \alpha}\cdot K, \label{treeNN}
\eeq
where $K$ is the factor due to the sum over soft gauge bosons produced
by instantons.  At low energies $K$ is known to grow exponentially as,
\beq
 K\sim  e^{ +c {2 \pi \over \alpha}({E\over E_{sp}})^{4/3}};\qquad c=O(1)
>0.
\label{Klow}\eeq
The difference between Eq.(\ref{tree2N}) and Eq.(\ref{treeNN}) reflects
the fact that in the latter there is a  sum over both the initial and
final (multiparticle) states while in the former the summation
is only over
the final states.

 It follows from Eq.(\ref{multi}) that the unitary
amplitude is limited by \beq A_{2 \to N} = {B K^{1/2} \over 1 + B^2 K^2}
\le {\rm const.}\,B^{1/2},\qquad B \equiv  e^{-2\pi/\alpha },
\label{ineq}\eeq
 hence
\beq
\sigma_{\Bvio} \sim |A_{2 \to N}|^2
               \le \exp - {2 \pi \over \alpha},
\label{half}\eeq
the  well-known "half-suppression" result.   The terminology is due to the
fact
that at the tree level  $\sigma_{\Bvio}
               \sim \exp - {4 \pi \over \alpha.}$

Note that this bound (Eq.(\ref{ineq}), Eq.(\ref{half}))
 is independent of the way $K$ depends on the energy,
i.e., does not depend on the low-energy approximation, Eq.(\ref{Klow}).
It is  thus  probably of a broader validity than it might first appear to be.
On the other hand, there is no guarantee that the upper bound is
actually reached: the correct statement is that $\sigma_{\Bvio}$ is
suppressed {\it at least} by the half (or more properly, the square root)
 of the standard 't Hooft factor.

If one does use the leading semi-classical result, Eq.(\ref{Klow}),
a somewhat stronger result follows. Namely,
 the full amplitude
would be dominated by the multiinstanton terms in Eq.(\ref{multi}),
as soon as the energy reaches a critical value such that
$A_{N \to N}^{tree} \simeq 1$ (it turns out to correspond to the value
$E_{max} \simeq 0.95 \sp$, in the leading semi-classical
approximation: see Eq.(\ref{Emax})).  But at this energy the tree
amplitude  $A_{2 \to N}^{tree}$ is still exponentially small!

It would  appear  then that the multi-instanton corrections  overwhelm the
single instanton contribution at an energy such that the latter is still
exponentially small,  invalidating  any argument based on
single instanton calculation extrapolated
to higher energies.  This scenario was termed  "premature unitarization"
by Maggiore and Shifman\cite{shif91}.

At present\cite{shif93}\daa\cite{kkr92b},
 there is no rigorous proof of premature unitarization, as
formulated above.  The estimate Eq.({\ref{Klow}), based on the leading
semi-classical approximation,  may receive important
modifications when higher order corrections are taken into account.
However, the alternative possibility - that the multi-instanton
contributions never overcome the single instanton term -  means that
 the $K$ factor appearing in Eq.(\ref{multi}), Eq.(\ref{ineq}) is never
 able to compensate the original t'Hooft suppression factor. In other
 words, the single instanton calculation may be saved, but it must stay
 exponentially small, after all!

 This discussion illustrates the essence of the arguments based on unitarity.
 Independent of the details (for instance, whether or not premature
 unitarization holds true),
 multi-instanton unitarization implies
the exponential (at least "half-") suppression of $\sigma_{\Bvio}$.

 A similar conclusion, in fact, follows also from a purely Minkowski picture
of our processes: the sphaleron resonance formation and its
decay\cite{manton}\co\cite{aoya87}\co\cite{mclerd87}.

Consider the Breit-Wigner formula for the $2 \to N $ amplitude \beq A_{2
\to N} \simeq { \Gamma_2^{1/2} \Gamma_N^{1/2} \over (E - \sp)+ i
\Gamma_{tot} }, \label{B-W} \eeq where $\Gamma_{tot} \simeq \Gamma_N$ is
the total decay width, $\Gamma_2 $ is the partial width to the two
particle
state. Eq.(\ref{B-W}) satisfies unitarity, ${\rm Im}A_{2\to 2} = \sum_N
|A_{2 \to N}|^2.$ Now, the sphaleron is a classical field with spatial
dimension $\sim 1/gv, $ hence the decay particles will have momenta of the
order of $ \sim gv $.
It follows that the average number of the secondaries is ${\bar N} \sim
{\sp / gv}\sim 1/\alpha$  (recall $\sp \sim  v /g.$)
  If one assumes a Poisson distribution for the number of the
secondaries - as is appropriate for a coherent state - then one ends up
with the estimate
 \beq |A_{2 \to N}|^2 \le \Gamma_2/\Gamma_{tot} \sim e^{-
{\bar N}} \sim e^{ - {\rm const.}/\alpha}. \eeq
 Thus the cross section for
the sphaleron formation is small because the latter is coupled weakly to
the two-particle initial state.

The fact that we find from the sphaleron picture an estimate similar
to the one obtained in the multi-instanton picture, is not really
surprising.  The sphaleron is an unstable state staying on the top
of the barrier separating two adjacent vacua of the SU(2) gauge
theory\cite{manton},
decaying with equal probabilities to $\Bvio$ and $\Delta (B+L) =0$
final states.   In the context of Euclidean, instanton
description
of $\Bvio$ transition, the sphaleron should correspond to multi (infinite)
instanton configurations.

Finally, it is possible to understand both types of arguments
based only on a very general feature of unitarity and
the particular aspect of involved dynamics - dominantly multi-particle
production.

 Consider\cite{konishi91}
 the s-channel unitarity equation for the forward elastic
amplitude  $f_1f_2 \to  f_1 f_2 \,\,(A_{el} \equiv A_{2 \to 2})$  ,
\bea
 {\rm Im} A_{el} &=& \sum_N |A_{2 \to N}|^2 \non \\
            &=& |A_{el}|^2 + \sum_{N \ne 2} |A_{2 \to N}|^2
\non \\
            &\propto&  \sigma_{el} + \sigma_{inel},
\label{sunit}\eea
where we assumed a simple form of unitarity appropriate for the
S-wave\cite{eden},
that
is adequate for processes described by a single  instanton or sphaleron, and
neglected all complications arising from spin.   As is well known, it leads
to the upper limit for an S-wave amplitude,
\beq
|A_{el}| \le 1.
\label{Swave}\eeq
The question is whether such a limit can be actually saturated
in the case of instanton-induced multi W boson productions.

It should be recalled that in the case of   high energy hadron
hadron scattering (due to strong interactions), more and more partial
waves come into play as the energy  increases, and this leads to a
much less stringent unitarity limit, the  so-called
 Froissart bound\cite{Froi} for the total cross section,
\beq  \sigma _{tot} \propto  \log^2 s. \eeq
(which amounts to $ |A_{el}(s, 0)| \le  s \log^2 s.$ )

Whether the limit (\ref{Swave}) can be reached, depends on
the dynamics, even if the reaction proceeds indeed through the S-wave only.
     Let us define the inelasticity $r$,
\beq
 r \equiv \sigma_{inel} / \sigma_{el} =
 \sum_{N \ne 2} |A_{2 \to N}|^2/|A_{el}|^2
\label{inelas}\eeq
Eq.(\ref{sunit}) and Eq.(\ref{inelas}) together imply a more stringent
bound,
\beq
\sigma_{\Bvio} \le \sigma_{tot} \simeq {1 \over s} {\rm Im}A_{el}
\le {1 \over s} {1 \over 1 + r}.
\eeq
We see that  whatever mechanism (in  pure S-wave)  gives $r \gg 1$
would lead to $\sigma_{\Bvio}$ much smaller than the geometrical
value ($\sim {1 \over s}$).

In a  more careful treatment one should  distinguish
 the baryon number
violating from baryon number conserving intermediate states
in Eq.(\ref{sunit}).  The argument essentially goes through without
modification, though, the reason being that the particular mechanism under
study
- multi-instanton sum - will give the same contribution to $\Bvio$ and
$\Delta (B+L) =0 $ processes, at its maximum (see
the next subsection).

The s-channel iteration of instanton anti-instanton chain is a crude
approximation which ensures the s-channel unitarity, Eq.(\ref{sunit}).
A four point amplitude must satisfy the t-channel (as well as the u-channel)
unitarity\cite{eden} also.  That  requires at least the instanton chain
 iterated also in the t- or u-channel.  Such contributions can  modify
 the S-wave dominance of the amplitude: cannot the electroweak high energy
 scattering then become of multi-peripheral type at high energies,
  with  a much
larger cross section, and become similar to the "soft" hadronic processes?

Such a possibility  cannot in principle be  excluded.  Nevertheless,
there is a difference here as compared to the multi-instanton
contribution iterated in the direct (s-) channel.  The large
 energy (large s)
 carried by the
initial particles  is  converted  into the energy  of  multiple gauge or
Higgs bosons. The effect of the final state summation is to enhance
drammatically the
leading  instanton amplitude\cite{ring90}\co\cite{esp90}.
 The effect is so large  as to lead to the violation of
 s-channel unitarity: one is
 {\it forced}
to take into account  higher order (multi-instanton) contributions, iterated in
the s-channel.   On the contrary,
no  reasons are known to suspect that multi-instantons  iterated in the
t-channel (for instance) are of any particular importance.  Multi-instanton
chains iterated in the $t$ or  $u$ channel presumably  remain negligibly small.

\subsection {Resummation of multi-instanton contributions}
These arguments above, leading to the "half-suppresion", based on a very rough
estimate, Eq.(\ref{multi}),  will be corroborated below
by  actually resumming
the multi-instanton contributions.  The treatment closely follows that  by
Musso and one of the autors\cite{konishi91}.

The  starting point will be the  one-instanton $R$-term calculation of the
$\Bvio$ cross section, or the imaginary part of the forward elastic
amplitude, reviewed in Section (1.3).  To generalize  the formula
to multi-instanton cases  one  needs   the knowledge of  the
{\it amplitude itself}  rather than  its imaginary part, since the former (more
precisely the S-matrix element) can be iterated  in the sense of Feynman
diagrams.

To compute the amplitude,
 we recall that if the instantons and
anti-instantons are far apart,  the single instanton S-matrix elements
can be
treated as  an effective Lagrangian, $ L_{inst} $. The  result
Eq.(\ref{Rformula}), Eq.(\ref{expW}), Eq.(\ref{expW2})
just corresponds to the
lowest contribution in $L_{inst} $ to the amplitude of W-boson production,
squared and summed over the final states.
Thus to get the amplitude,  one  replaces the "cut propagator"
\beq
 \pi \int {d{\bf k}\over \omega_k } =
  2 \pi \int d^4k \,\theta (k^0) \delta(k^4)
\eeq
appearing in Eq.(\ref{expW}) by the uncut
propagator
\beq
\int {d^4k\,i \over {k^2+i\epsilon}}.
\eeq
One finds thus,
\beq
<{\bf p},-{\bf p}| A | {\bf p},-{\bf p} > =
i\, e^{-16 \pi^2 / g^2} \int dt \,d{\bf x} \, d\rho \, d\rho^{\prime}\,
C(\rho,\xi,\rho^{\prime},\xi^{\prime})\, e^{W^{\prime}}.
\label{amplitude}\eeq
The exponent $W^{\prime}$ is given by
\beq
W^{\prime}=-iEt-\pi^2v^2 (\rho^2 + \rho^{\prime 2}) +
 {\pi \over g^2}\rho^2
\rho^{\prime 2} \int {d^4 k \, i\over k^2 + i \epsilon} e^{ikx}
(3k_0^2 + {\bf k}^2).
\label{Wamplitude}\eeq
 Now, since
\beq
\int {d^4 k \, i\over k^2 + i \epsilon} e^{ikx} =- {4\pi^2\over
x^2-i\epsilon},
\eeq
the integral in $k$ in $W^{\prime}$ is equal to
\beq
{32\pi^2(3x_0^2+{\bf x}^2)\over (x^2-i\epsilon)^3}.
\eeq
Computing the ${\bf x}$ integration by the saddle point method, one
finds at the saddle point (${\bf x} =0$),
\beq
W^{\prime}=-iEt-\pi^2v^2 (\rho^2 + \rho^{\prime 2}) + {96\pi^2
\rho^2 \rho^{\prime 2} \over g^2 t^4} = W:
\label{Wprime}\eeq
namely, in the leading order we find the same exponent for the forward
amplitude as for its imaginary part. This is consistent with the fact that
the  leading
  $ i - a $ contribution to the  forward amplitude is purely
imaginary.
 Computing the resulting integrals over $t$ and $\rho, \rho^{\prime} $ by
the saddle point method as before , one finds thus
\beq
<{\bf p},-{\bf p} | A | {\bf p},-{\bf p} >
 =  i\, \exp \Bigl(-{16\pi^2\over g^2} + 3 \bigl({3E^4\over
8\pi^2g^2v^4} \bigr)^{1/3}\Bigr),
\eeq
in accordance
with Eq.(\ref{growth}).

\smallskip
The contribution from $n$ pairs of $i - a $  (Fig. 10) to the forward
amplitude
 can  be written down by generalizing Eq.(\ref{amplitude}),
Eq.(\ref{Wamplitude}). If we neglect various
pre-exponential factors, it reads,
\beq\label{sono}
 <{\bf p},-{\bf p} | A^{(2n)} | {\bf p},-{\bf p} >
   = i^{2n-1} e^{-16 \pi^2 n / g^2} \int ... \int \prod_{i=1}^{2n-1}
dt_i \prod_{i=1}^{2n} d \rho_i \exp W.
\eeq
\beq
W=-iE\sum_{i=1}^{2n-1} t_i - \pi^2 v^2 \sum_{i=1}^{2n} \rho_i^2
+\Cost \sum_{i=1}^{2n-1} \rho_i^2 \rho_{i+1}^2/ t_i^4,
\label{Wmulti}\eeq
In arriving at Eq.(\ref{Wmulti}) the integrations over the relative
instanton
orientations have been done by the saddle point approximation which
effectively set all instantons  and anti-instantons
 alligned in the $SU(2)$ space. The
integration over the relative instanton (space) positions yields
\beq\label{stufo}
{\bf x}_1= {\bf x}_2 = ...= {\bf x}_{2n-1} = 0.
\eeq
The remaining   integrations over the
relative instanton time coordinates $t_i$, and their sizes $\rho_i$,  can also
be done easily.  See   Appendix A.

The resulting 2n instanton contribution to
the forward $ff \rightarrow ff$ amplitude is now
(in terms of the dimensionless variable $x\equiv E/\sp$)
\beq
A^{(2n)} \propto  i^{2n-1}H^{2n-1}
e^{
{-16\pi^2\over g^2}
\left( (1-1.0817 x^{4/3})n+0.7672 x^{4/3}\right)
},
\label{A2n}\eeq
where the real factor  $H$ contains the power
 contribution from the fermion zero modes
in the alternative  $i - a $ "bonds" as well as the preexponential
factors arising from the gaussian integrations.

Note that the phase factor $i^{2n-1}$ in $A^{(2n)}$ originates from
the continuation from
2n $i - a$ centers in the Euclidean space to Minkowski spacetime positions.
(One missing $i$ is due to the standard definition of the amplitude.)  All
other  Gaussian integrations over $t_i$,s and $\rho_i,s $ are real.
As a result the sign alternates in the sum over $n$ (see Eq.(\ref{multi})),
 and gives
\beq
A \equiv
\sum_{n=1}^{\infty} A^{(2n)}
 \simeq
 i {H\,e^{{-16\pi^2\over g^2}[1-1.0817 x^{4/3}]} \over 1 +
H^2\, e^{{-16\pi^2\over g^2}[1-1.0817 x^{4/3}]} }
e^{-{16\pi^2 \over g^2}[0.7672 x^{4/3}]}.
 \label{resummed} \eeq

This concludes the calculation of the forward elastic amplitude
in the leading semi-classical n-instanton approximation,
resummed over n.

 \smallskip
There are several noteworthy features in Eq.(\ref{resummed}).

\noindent (i) As expected, the resummed amplitude behaves
in a way qualitatively different from the single instanton
contribution ($n=1$ term only).
The resummed amplitude, which is unitary, turns out to be
exponentially suppressed at all energies.  It  reaches the maximum
\beq
|A|_{max} \simeq  \exp -{2\pi \over \alpha}(1.333)
\eeq
at
\beq
E_{max}\simeq 0.95 \sp.
\label{Emax} \eeq
where $1 - 1.0817 x^{4/3} \simeq 0$.
At higher energies it is exponentially damped.
The suppression Eq.(\ref{Emax}) is somewhat stronger
 than the naive "half suppression"
factor $\exp(-2\pi/\alpha)$, in agreement with our general conclusion.

\smallskip
\noindent  (ii) That the resummed amplitude
 $A \equiv \sum_{n=1}^{\infty} A^{(2n)} $ satisfies the s-channel
unitarity,
\beq
(A_{2 \rightarrow 2} - A_{2 \rightarrow 2} ^*)/2i =
 \sum_N A_{2 \rightarrow N} \times A_{2 \rightarrow
N}^* ,
\label{unitarity} \eeq
can be seen as follows. Consider
 $A_{2 \rightarrow 2}$  and $ A_{2 \rightarrow N}, $ both computed
 in the  leading 2n-instanton  approximation,
summed over n.  Compare the two sides of the unitarity equation, namely
$ Im  A_{2 \rightarrow 2}$ and  $\sum_N | A_{2 \rightarrow N} |^2.$
The latter can be written as
\beq \sum_N \sum_{l,m=1}^{\infty} A^{(l)}_{2 \rightarrow N}
(A^{(m)}_{2 \rightarrow N})^* = \sum_N
\sum_{n=1}^{\infty}\sum_{l=1}^{2n-1}A^{(2n-l)}_{2 \rightarrow N}
(A^{(l)}_{2 \rightarrow N})^*. \label{unitt}\eeq

Recall   that the sum over the intermediate states $N$ (corresponding to
cut propagators)
 gives rise to the same exponent
 as the sum over virtual intermediate
 states (with
uncut propagators),  as  noted before (after Eq.(\ref{Wprime})). Taking
into account the appropriate phase factor  the $(n,l)$ term of
 Eq.(\ref{unitt}) is
\beq\sum_N A^{(2n-l)}_{2 \rightarrow N}
(A^{(l)}_{2 \rightarrow N})^* = i^{2n-l-1} (-i)^{l-1}
 |A^{(2n)}_{2 \rightarrow 2}|. \eeq
These terms are seen to be in one-to-one correspondence with various
cuts of multi-instanton chain for $A^{(2n)}_{2 \to 2}$.  Indeed, the
identity, \beq
\sum_{l=1}^{2n-1} i^{2n-l-1} (-i)^{l-1} = (-)^{n-1}
= (i^{2n-1} - (-i)^{2n-1})/2i \label{cuts}  \eeq
shows
 that Eq.(\ref{unitarity}) is satisfied in this approximation.

\smallskip
\noindent (iii) Note that both types of
cuts corresponding to $\Delta (B+L) \ne 0 $
and  $\Delta (B+L) =0 $ intermediate states
 appear in the unitarity equation.  However,
 it is not difficult  to separate the two types of contributions
in Eq.(\ref{resummed}), once one  identifies the contributions of various
cuts, Eq.(\ref{cuts}).  One finds
\beq
{\rm Im}\,A=Disc_{\Delta (B+L) \ne 0} A +Disc_{\Delta (B+L) =0} A.
\eeq
\beq\label{disc1}
Disc_{\Delta (B+L) \ne 0}\,A=
{H\,e^{{-16\pi^2\over g^2}[1-1.0817 x^{4/3}]} \over
\bigl(1 +
H^2\, e^{{-16\pi^2\over g^2}[1-1.0817 x^{4/3}]}\bigr)^2 }
e^{-{16\pi^2 \over g^2}[0.7672 x^{4/3}]},
\eeq
\beq\label{disc2}
Disc_{\Delta (B+L) = 0}\,A=
{H^3\,e^{{-32\pi^2\over g^2}[1-1.0817 x^{4/3}]} \over
\bigl(1 +
H^2\, e^{{-16\pi^2\over g^2}[1-1.0817 x^{4/3}]}\bigr)^2 }
e^{-{16\pi^2 \over g^2}[0.7672 x^{4/3}]},
\eeq
and
\beq
\sigma_{\Bvio} \propto
 ({1\over s}) Disc_{\Delta (B+L) \ne 0}\, A.
\eeq
Clearly all the point made above in (i) applies to $\sigma_{\Bvio}$
which is the quantity one is really interested in.

Finally let us note that at the energy Eq.(\ref{Emax}) at which
$\sigma_{\Bvio}$ takes the maximum value (and where
 $1 - 1.0817 x^{4/3} \simeq 0$), the baryon number violating and
conserving cross sections are the same:
\beq
\sigma_{\Bvio} \simeq \sigma_{\Delta (B+L) =0}
\eeq
to  exponential accuracy,
in accordance with the sphaleron resonance formation picture.

\smallskip
\noindent (iv) Instead of $2 \to N$ amplitude considered above,
one can study $N \to N^{\prime}$, with the coherent states method
of Khlebnikov et al\cite{krt}.    Consider in particular the forward
amplitude,
\beq
ff+N {\rm \; gauge \; bosons} \longrightarrow  ff + N {\rm \;
gauge \;
 bosons},
\eeq
in the multi-instanton approximation, summed over $N.$
The R-term technique for
 doing the sum over the number of particles in the
intermediate states used above can be easily extended to the initial
state (see also Section 2.2).

One finds that
\beq
 |A_{ff+N  \rightarrow ff+N}
|_{N=N_{max}}\simeq  |\sum_{all\,N} A_{ff+ N \rightarrow
ff+ N}|
\eeq
for some $N_{max}$ of order of
 $O\bigl(({1\over \alpha})({E\over \sp})^{4/3}\bigr)$
by using the well known relation between a coherent state
and a state with a given number of particles.
Also,
the right hand side of this equation can be computed just as in the
case of the forward $2 \to 2$ amplitude, Eq.(\ref{resummed}).
 The crucial difference is however that now  no "end point effect"
 is present and as a result one gets\cite{konishi91}
\beq
A_{ff+N {\rm \; gauge \; bosons} \rightarrow ff+N {\rm \;  gauge \;
 bosons}}|_{N=N_{max}}
 =
 i {H\,e^{{-16\pi^2\over g^2}[1-1.0817 x^{4/3}]} \over 1 +
H^2\,  e^{{-16\pi^2\over g^2}[1-1.0817 x^{4/3}]} }.
\label{nosupp}
\eeq
a result similar to Eq.(\ref{resummed}) but without the last exponential
factor.
Thus the many-to-many amplitude would reach the
unitarity limit,  $\exp - {4\pi \over \alpha}\cdot 0 $, in contrast to
the $2 \to \all$ amplitude.

\subsection{High temperature or
 high density transition as compared to processes at  high energies}

As is well known\cite{hight},
high-temperature $\Bvio$ transitions occurs classically
(and without barrier penetration factor)
in the standard model, if the temperature is above the energy barrier,
that is if \beq
T \ge \sp
\eeq
Analogously, the ground state with finite fermion density will decay
without any suppression if the  Fermi sea level is sufficiently high,
i.e., of the order of the sphaleron mass\cite{highd}.

That a similar disappearance of the tunnelling factor might occur also
in {\it high-energy} scattering,  was the original motivation for the
earliest works \cite{ring90}\co\cite{esp90}.  However, there is a clear
physics difference here.
  A characteristic feature of the scattering  processes is that the high
   energy of the initial channel is carried by just two energetic
   particles.  In the high temperature (or high density) transitions,
   on the contrary, it is
   expected that the relevant initial states,  at  a given (high) energy
   and with a given set of quantum numbers,  are mainly
    multiparticle states. For
    there are many more such states as compared to two particle
   states, hence they will dominate the process for purely statistical
   reasons.
 Thus although   many-to-many  amplitudes might well become
   unsuppressed at the sphaleron energy (see  Eq.(\ref{nosupp})),
   the two-to-many amplitude will remain small.
A suggestion put
forward by Diakonov and Petrov\cite{dp}\co\cite{ring92},
 that the decay rate of
the state with    a finite density (with the Fermi level $\mu $) is
proportional to $\Bvio$ cross section (with $E \sim \mu $), seems to be
invalidated, once this difference in the initial states are taken into
account\cite{smilga}.



\section {Fermions in the valley}
In this section the whole problem will be analyzed from a
somewhat different point of
 view.  The behavior of the chiral fermions in the valley
background is studied, and the $\Bvio$ cross section is analysed  by using the
optical  theorem.

\subsection {Unitarity puzzle and Spectrum of the Dirac operator}
As already noted in Sections (1.3) and (4.1),
 the total $\Bvio$ cross section
can be computed  via unitarity, i.e., as an appropriate part of the
imaginary part of the forward elastic amplitude,
\beq
1 + 2  \to  1 + 2.
\label{forward}\eeq
However, a  problem concerning unitarity and chiral anomaly
arises, which can be formulated as follows\cite{guida}.
 The optical theorem states that the cross section,
\beq
1 + 2 \to X,
\label{production}\eeq
summed over all possible $X$, is equal,
 apart from a kinematical factor, to the imaginary
part of the forward elastic amplitude, Eq.(\ref{forward}).
Now consider a particular class of processes (\ref{production})
 induced by an $SU(2)$ instanton,
with
\beq
  \Bvio; \qquad \Delta (B - L) =0.
\label{selection}\eeq
Sum over the final states satisfying (\ref{selection})
 should give  {\it a part}
 of the full  imaginary
part of the elastic amplitude:
\beq
{\it Anom}\, {\it Im}\,  A_{2 \to 2}  =  \sum_{\Bvio} |A_{2 \to X}|^2,
\label{anouni}\eeq
see Eq.(\ref{disc1}-\ref{disc2}).
Now, for an (anti-) instanton background,
which is relevant for the calculation
of the right hand side of Eq.(\ref{anouni}),
 each right (left) handed fermion field
has a zero mode. The standard functional integration over fermions yields  a
product of these zero modes; by going to momentum space and by applying
the LSZ amputation one finds the S-matrix elements consistent with the
instanton selection rule
Eq.(\ref{selection}).

How to calculate  the left hand side of Eq.(\ref{anouni})?  Being a part
of the elastic amplitude, it
 must arise from a four point function computed in a
 background, topologically (globally)
equivalent to the trivial, perturbative vacuum.  To be equal to the total
anomalous ($\Bvio$) cross section, however,  such a background must have a
nontrivial topological structure,  for instance,  similar to
a widely separated instanton
anti-instanton pair.  More precisely, the work of Arnold and
Mattis\cite{mat91b}
suggests that one should consider something like the  valley.

The problem is  that no fermion zero modes exist in the valley
background\cite{guida}
(see Appendix B for a sketch of the proof).  The spectrum
of the Dirac operator in the valley background Eq.(\ref{valley})
 is proven\cite{guida} to be
the same as that of the free Dirac operator, with the continuum spectrum
$(-\infty, \infty)$ and with no normalizable  zero or nonzero modes.
Thus one wonders how the left hand side of Eq.(\ref{anouni}) can be
computed, which should somehow be dominated by the standard  lefthanded
or righthanded zero modes, in order to match the right hand side.

Note that this "unitarity puzzle" by no means depends on the use
of the fermions
as external particles in an essential manner, although the puzzle looks
much  neater with fermions.  In the
literature,  often  theories without fermions are considered, on the basis
that the main
problem lies in topologically nontrivial aspects of {\it gauge} field
 dynamics (which is quite true).   The use of gauge bosons
as external particles however does not eliminate the unitarity puzzle.
For instance, the validity of the leading order "semiclassical"
approximation (\ie
substituting the instanton solution in the external lines) is not at all
 obvious for the left hand side of Eq.(\ref{anouni})
(especially in  the regime of strongly overlapping instantons),  if it is more
reasonable for the production amplitude appearing on the right hand side.

To solve the "unitarity puzzle", one must study the elastic amplitude
starting from the four point function computed in an appropriate
 background of
instanton  anti-instanton type (which will be  approximated here by the
valley),
and must show that the imaginary part of the forward elastic amplitude indeed
contains an anomalous piece which reproduces the right hand side of
Eq.(\ref{anouni}).  The work reviewed in the next section is a first step
towards such a solution.

\subsection{Fermion Green function in the valley:
widely separated instanton
anti-instanton pair}

The four point function,
\bea
<T\psi_1(x) \psi_2(u){\bar \psi}_1(y) {\bar \psi}_2(v)>^{(A_{valley})}
\non \\
=\int{\cal D}\psi{\cal D}{\bar \psi} \, \psi_1(x) \psi_2(u) {\bar \psi}_1(y)
 {\bar \psi}_2(v)
\,{\rm e}^{-S} /Z^{(A=0)};
\eea
\beq
S=\sum_{j=1}^{N_F} \int d^4x\, i\,{\bar \psi}_j {\bar D}\psi_j
\eeq
in the fixed background of Eq.(\ref{valley}), has been studied by two of
the present authors\cite{guida}.
As the functional integral factorizes in flavour the quantity of
interest
are (suppressing
the flavour index),
\beq
I(x,y) = \int {\cal D}\psi{\cal D}{\bar \psi}\, \psi(x) {\bar \psi}(y)
e^{ -\!\int \! d^4x\, i\,{\bar \psi}{\bar D}\psi},
\eeq
and
\beq
{\cal Z}=  \int {\cal D}\psi{\cal D}{\bar \psi}
e^{ -\!\int\! d^4x \,i\,{\bar \psi}{\bar D}\psi} = \det \dbar,
\eeq
where it is assumed that $\det \dbar$ is suitably regularized.

The key point in this analysis  is to  introduce complete sets of orthonormal
modes $\{ \eta^{(a)}_n \}$
and $\{{\bar \zeta}^{(i)}_n \}$,  $n=0,1,2,....$,
  for the left-handed and right-handed
fermions, respectively.
  They are eigenstates of $ D^{(a)} \dabar$
 and $\dibar D^{(i)} $:
$$  \dabar \etm = {\bar k_m} \zema\,\, (m=0,1,...),
  \qquad  D^{(a)} \zema = k_m \etm\,\,  (m=1,2,...),       $$
  \beq
  {\bar D}^{(i)} \emi =  l_m \zetm\,\, (m=1,2,...),
  \qquad
D^{(i)} \zetm = {\bar l_m} \emi\,\,  (m=0,1,...),
\eeq
where
\beq
{\bar k_0}={\bar l_0}=0.
\eeq
 The covariant derivatives $D^{(a)},\, D^{(i)}$ are defined
with respect to  the anti\-instanton  and
instanton parts of Eq.(\ref{valley}).  Accordingly
the zero modes are those in the regular gauge (for the lefthanded mode)
and in the singular gauge (for the righthanded one), respectively.

The functional integration can  then  be defined as:
$$
\int{\cal D}\psi{\cal D}{\bar \psi} \equiv \prod_{m,n=0} da_m\,
d{\bar b}_n;
$$
\beq
\psi(x)  = \sum_{m=0}^{\infty} a_m \eta^{(a)}_m(x),\qquad
{\bar \psi}(x) = \sum_{n=0}^{\infty} {\bar b}_n {\bar \zeta}^{(i)*}_n(x).
\eeq

As is clear from the way these equations are written,
  the system is first put in a large but finite  box
of linear size $L$
( such that $ L \gg  R, \rho $)
so that all modes are discrete.  After the derivation of  Eq.(\ref{result})
below
(i.e., after the sum over the complete sets is done),  however,
$L$  can be sent to infinity without any difficulty.

The two point function $ I(x,y)  $ can be written as
\bea \label{twopo}
  I(x,y) &=& \det{\bar D}\, \langle x|\dinv|y \rangle  \\
      &=&\det {\bar D}\, \{ \langle x|a,0 \rangle
\langle a,0|\dinv |i,0\rangle \langle i,0|y \rangle
    +\sum_{m\ne 0}\langle x|a,m\rangle
\langle a,m|\dinv|i,0\rangle \langle i,0|y \rangle  \non \\
     &+&\sum_{n\ne 0}\langle x|a,0 \rangle
\langle a,0|\dinv|i,n\rangle \langle i,n|y \rangle
      + \sum_{m,n\ne 0}\langle x|a,m\rangle
\langle a,m|\dinv|i,n\rangle \langle i,n|y \rangle \}: \non
\eea
the term proportional to the product of the zero modes
 has been singled out.   We wish to compute $I(x,y)$ at small
$\rho/R$. To do this, first
let us write
\beq
 \dbar = \pmatrix{d&v_1&\ldots&v_n&\ldots\cr
                         w_1&X_{11}&\ldots&X_{1n}&\ldots\cr
                 \vdots&\vdots&\ddots&\vdots&\ddots\cr
                   w_m&X_{m1}&\ldots&X_{mn}&\ldots\cr
                  \vdots&\vdots&\ddots&\vdots&\ddots. \cr}
\eeq
The idea is that the matrix elements involving either of the zero modes,
$d,\, v_n,\, w_m, $ are all small by some overlap intergrals while the
matrix elements $X_{mn}$ are large because the wave functions
 of non zero modes
are extended to  all over the spacetime.

The inverse matrix $\dinv$ is given by:
\bea
  (\dinv)_{00} &=& 1/( d - vX^{-1}w ) \non \\
 &=& d^{-1} + d^{-2}v_m (X^{-1})_{mn} w_n + \cdots; \non \\
(\dinv)_{mn} &= &(X - {1\over d}w \otimes v) ^{-1}=
X^{-1} ( 1 - {1\over d}w \otimes v X^{-1} )^{-1} \non \\
 &=& (X^{-1})_{mn} + d^{-1}(X^{-1})_{ml} w_l v_k (X^{-1})_{kn}
        + \cdots, \non \\
 (\dinv)_{0n}& =& - d^{-1} v_l (\dinv)_{ln},  \non \\
(\dinv)_{m0} &= &- (\dinv)_{00} X^{-1}_{mk} w_k,
\label{invers}\eea
where $X^{-1}$ is the inverse of the submatrix $X$ in the space orthogonal
to the zero modes.

Inserting Eq.(\ref{invers}) into Eq.(\ref{twopo})
 and after some algebra  one finds a remarkably simple (and still exact)
 expression for $I(x,y)$:
 \bea
 I(x,y) &= \det X\,\{\bra x|a,0\ket -
 \bra x | X^{-1}\cbar|a,0\ket \}\, \{\bra i,0|y\ket -
\bra i,0|\bbar X^{-1} |y \ket \}   \non \\
 & + \det \dbar \, \bra x | X^{-1} | y \ket,
 \label{result}\eea
where  $\cbar \equiv C_{\mu} {\bar \sigma}_{\mu};\,\,
\bbar \equiv B_{\mu} {\bar \sigma}_{\mu}$ in Eq.(\ref{result})
 are defined by:
\beq
D_{\mu}^{(valley)} = D_{\mu}^{(a)} + C_{\mu} = D_{\mu}^{(i)} + B_{\mu}.
\eeq
The function $C_{\mu}$ is the (modified) instanton field while  $B_{\mu}$
is the (modified) anti-instanton.

Eq.(\ref{result}) displays nicely the main features of the two point
function in the valley background.  The effect due to the zero modes is
separated and everything else is expressed by the smoother two point
function,
\beq
S^{\prime}_{x,y} = \bra x| X^{-1} |y \ket.
\eeq

 $\sprime$ can be assumed to behave
 at large $x$ and $y$ (with $x_i$ and $x_a$ fixed)
 as
\bea
 \sprime & \sim & U^{\dagger}(x) S_F(x,y)  U(y), \non \\
   U(x) & =& { {\bar \sigma}_{\mu} (x-x_a)_{\mu}
\over \sqrt{(x-x_a)^2}}, \label{ipotesi}
\eea
 where $S_F$ is the free Feynman propagator. This behavior is suggested by the
 fact that the valley field has a pure gauge form at large x,
\beq   A_{\mu}^{(valley)} \sim {i\over g}\, U^{\dagger}\partial_{\mu}\,U
   \sim O({1\over x}).   \label{3_10} \eeq
To proceed further one  assumes that
\beq \det X / \det {\bar \partial}  = {\rm const.}
  \label{3_10bis}\eeq
as $\rho/R \to 0.$
Next the ratio  $\det \dbar / \det X $ can be estimated as follows
(see (\ref{invers})):
\beq {\det \dbar \over  \det X} = ((\dinv)_{00})^{-1} \simeq d
\simeq {\rm const.}\, \rho^2/R^3,   \label{3_10ter}\eeq
where use was made of
$$ d =  \dbar_{00}= \bra i,0|\cbar| a,0 \ket =
 \int_z \zet0\!(z)^*  \cbar\!(z)\, \et0\!(z)
   \sim \rho^2/R^3. $$

\noindent
Combining Eq.(\ref{3_10bis}) and Eq.(\ref{3_10ter}) gives
\beq
{\det \dbar \over  \det {\bar \partial}} \sim  \rho^2/R^3. \label{3_10quater}
\eeq

 With Eq.(\ref{3_10quater})
and Eq.(\ref{ipotesi}) in Eq.(\ref{result}) one can estimate the amplitude
and the leading contribution to its anomalous part.

There is a  subtlety here.  Because the background (in
the gauge of Eq.(\ref{valley})) reduces asymptotically to a pure gauge
form, Eq.(\ref{3_10}), and does not vanish sufficiently fast,
 the standard LSZ procedure cannot be used to
extract the amplitude.  The correct procedure is to go to a more
 physical gauge
\bea
 {\tilde A}_{\mu}^{(valley)} &= &
U ( A_{\mu}^{(valley)} + { i\over g}
\partial_{\mu} ) U^{\dagger},  \non \\
   U(x) &= & { {\bar \sigma}_{\mu} (x-x_a)_{\mu}
\over \sqrt{(x-x_a)^2}},
\eea
by appropriately transforming the fermion four point function, before
the  standard LSZ procedure is applied\cite{guida}.

The contribution of the first term of Eq.(\ref{result}) the
elastic amplitude is found to be, to leading order in $\rho/R$,
proportional to
\beq
\rho^2  \exp(ip \cdot x_a) \exp(-iq \cdot x_i).
\eeq
 One  finds also a correction proportional to   $\rho/R $ times this
 factor,
coming from the terms containing the function $B$ and $C$ in
Eq.(\ref{result}).
These terms are precisely what one  has been looking for: the part of the
elastic amplitude, whose imaginary part  is going to match the right hand
side of the unitarity equation, Eq.(\ref{anouni}).

The second term of Eq.(\ref{result}), in contrast, mainly gives rise to
 non-anomalous discontinuity, associated with
 $\Delta (B+L) =0$
 intermediate states.

Thus at least for the valley corresponding to widely
separated  instanton anti-instanton configuration, the elastic amplitude
contains the piece which reproduces the square of the production amplitude
computed in the single instanton background.

The {\it general} recipe for calculating the left hand side of the "anomalous"
unitarity
relation,  Eq.(\ref{anouni}),  is however not yet known  to the best of
our knowledge.

\smallskip
We have thus far considered  the behavior of fermions
in the valley  which corresponds to a
particular kind   of instanton anti-instanton pair configuration,
oriented in the
maximally attractive direction\cite{yung}
 and with a particular interaction term\cite{kr91c}.
Such a field is supposed to be of particular relevance in  the problem of
baryon number violation, as reviewed in Section 1.2.

 A related problem of the
 fermion propagation in a background of {\it simple sum}  of widely
 separated  instantons and
anti-instantons (in singular gauge),
 was studied  earlier.    Such
 backgrounds  might be important in modelling some feature of
 the physical vacuum of
QCD\cite{gross}\co\cite{andgros}\co\cite{liquid}.
It is Lee and Bardeen\cite{Lee} who, in such a
context, showed for the first time
that the fermion propagator can be  in some sense dominated by
a term proportional to a product of  the standard zero modes
 (for widely separated
 instanton-anti-instantons), in spite of the absence of any fermion zero
 modes in the background considered.

 Also, the method used by Lee and Bardeen (they work with the
 equation satisfied by the fermion Green function, rather than
doing  the functional integration ) is very different with our own;
of course,  our formula reproduces theirs to the
leading order in $\rho/R$, if one uses  the simple sum of
 instanton and anti-instanton instead of the valley and sets
$X^{-1} \simeq  S_F + (S_a-S_F) + (S_i-S_F).$

\subsection{Overlapping instanton anti-instanton pair}

The valley background used above reduces to the vacuum field
at $R=0$\fnm{i}\fnt{i}{The triviality of the valley
for $R=0$ is a feature independent of the choice of
 the weight $w$\cite{kr91c}.}.
One wonders whether the transition to a purely perturbative field
occurs gradually and  only terminates at precisely $R/\rho =0$ , or
it takes place abruptly at a finite value of $R/\rho $, probably of order
of unity.

There is a strong indication that the latter possibility is realized. The
first indication\cite{guida}
 comes from the numerical and analytical study of the
integrated  topological
density,
\beq
 C(x_4) \equiv -\int_{-\infty}^{x_4} \int d^3x {g^2\over 16 \pi^2}
\hbox{\rm Tr }F_{\mu \nu}
 {\tilde F}_{\mu \nu} = {\cal N}_{CS}(x_4)-{\cal N}_{CS}(-\infty),
 \label{4_1} \eeq
as a function of $x_4$ for several values of $R/\rho$,
for the valley background of Eq.(\ref{valley}). (See Fig. 11.)
    ${\cal N}_{CS}(x_4)$ is
 the Chern Simons number
\beq {\cal N}_{CS}(x_4)\equiv -\int d^3 \!x {g^2\over 16 \pi^2}
\epsilon^{4ijk}\hbox{\rm Tr }(F_{ij}A_k-{2\over 3} A_i A_j A_k ).
\label{csn}\eeq
  The instanton and anti-instanton are
situated at $ ({\bf 0}, R/2) $ and at $ ({\bf 0}, -R/2) $, respectively.

It can be seen from Fig.11   that the topological structure
is well separated and localized at the two instanton centers
only at relatively large values of $R/\rho$,   $R/\rho \ge 10 $:
for such $R/\rho$,    $C(x_4)$
reproduces locally the situation of single instanton background
 (near $x_4= R/2$)
and that of anti-instanton (near  $x_4=-R/2$).
Vice versa, for
small $R/\rho \le 1 $  the gauge field is seen to
collapse to some insignificant fluctuation
around zero, not clearly distinguishable from ordinary perturbative ones.
(These statements are, admittedly, a little vague and not very precise.
See the next
subsection for  an attempt to make them more precise.)

In Fig. 12  is plotted  also the behavior of the maximum of each curve,
corresponding to $C(0)$, as a function of $R/\rho$.
 The behavior of $C(0)$ is powerlike both
at large and small R:
\beas
C(0) &\sim& 1 - 12(\rho/R)^4, \quad R/\rho \gg 1, \\
C(0) &\sim&\qquad  {3\over 4} (R/\rho)^2 ,\quad R/\rho \ll 1 .\eeas

(Actually,  the exact expression for $C(x_4)$ in terms
of $x_4$ can be found by using conformal transformations\cite{prov}).
In particular,
\beq
C(0)=3({z-1\over z+1})^2-2({z-1\over z+1})^3, \label{c0}\eeq
where $z$ is defined in (\ref{1_7}).)

The behavior of $C(0)$ supports  the idea that the transition to a
perturbative background  is a sharp one. In particular, the quadratic
 behavior of  $C(0)$  in $R/\rho $ found at small $R/\rho $
  is nothing but the reflection
of the perturbative quadratic fluctuation. This is particularly
transparent if one
uses the gauge\cite{guida} in which
 $$ A_\mu  \propto  R. $$

Altogether,
 Fig. 11 and Fig. 12 indicate that the instanton and anti\-instanton start
to melt at around $R/\rho \simeq 5 $  and go through the transition
quickly, the center of the transition  to a
purely perturbative regime  being at
 around $R/\rho =1$.  The known behavior of the valley field
 action\cite{verb}\co\cite{kr91c},
 which sharply drops from near the two instanton value $4\pi /\alpha $  to zero
 around the transition region, $R/\rho \simeq 1$,
(see Fig. 3)  is perfectly consistent
 with this conclusion.

\subsection{Level-crossing in the valley}

These  discussions, though convincing intuitively, do not provide  a
quantitative answer as to the value of $R/\rho $ at which such a
transition occurs.   The best thing to do would be to analyse the fermion
four point function directly around $R/\rho  \sim  1$,   which however
appears to be a difficult task for the time being.

The next-best thing to do is to study the level crossing of the
chiral fermion in the background of the valley, and see whether a
{\it  qualitative} change occurs at a finite value of $R/\rho $ .
The idea is the following.

As is well known,  one way to understand the anomalous chiral fermion
generation by a topologically non trivial gauge fields,
\beq
\Delta   N_R -  \Delta  N_L  =
\int d^4 x {1\over 16 \pi^2}F_{\mu \nu}
 {\tilde F}_{\mu \nu} = {\cal N}_{CS}(-\infty)-{\cal N}_{CS}(\infty),
 \eeq
is through the spectral flow of the eigenvalues of the Dirac Hamiltonian
$\H $, where $\H $ is defined by
\beq
 i\gamma _\mu D_\mu=i\gamma_0 (D_0 + \H) .
\eeq
   For instance, in the case of a single instanton,
one finds  (in the gauge  $A_0 =0$) that   $A_i( {\bf  x},  -\infty)$
and $A_i( {\bf  x},  \infty)$  are gauge equivalent: the spectrum of
$\, \H \, $ is the same at $t = \pm\infty$.   This however does not imply
that individual levels $E_j(t) $ are the same at  $t = \pm\infty$.   In
   fact,  the existence of  one right handed  zero
   mode (which is known to be there  from the index theorem\cite{as}),
   implies that  precisely one
   right handed mode  crosses zero in going from $t =-\infty$ to
$t =\infty$  so that (in adiabatic approximation\cite{alvarez}),
$$
 \psi _1 (t)  \sim e^{   - E_1(-\infty ) t}; \qquad   t \to -\infty $$
 \beq
 \psi _1 (t)  \sim e^{   - E_1(\infty ) t}; \qquad   t \to \infty
\eeq
with  $E_1(-\infty )  <0; \,\,E_1(\infty )  >0$.      Otherwise,  this
mode  would not be normalizable in four dimensions.

 Furthermore the point $t$ at which the level crossing occurs, is
 characterized by the fact that  the equation ${\cal H} \eta  = 0$
 has for such $t$  a solution, normalizable in  {\it three} space dimensions.

In the case of the valley background,   $A_i( {\bf  x},  -\infty)$
and $A_i( {\bf  x},  \infty)$  are not only gauge equivalent but
corresponds to the same Chern-Simons number (that is to say that
the full integrated topological density - the Pontryagin number -
is zero).   Thus we expect that   $E_j(-\infty)  = E_j(\infty) $ also for all
$j$.
Nevertheless, at least for widely separated instanton anti-instanton pair,
we expect from clustering argument  a non trivial spectral
flow  such as in Fig. 13a:
a right handed fermion level must cross zero at (more or less)
the instanton  time position  and then cross zero back  near the
anti-instanton  site.   In this way  the physics  of instanton  or
 of anti-instanton (anomalous fermion
generation or annihilation) would be locally
reproduced by the valley.

If  such a picture is confirmed  for large  $R/\rho $,
 then one can  ask what happens  for small $R/\rho $.    The strategy is thus
 to  check the method  of level crossing  where physics is well
 understood,  and then use the  same  tool  to  explore the
unknown territory.

The analysis made by the present authors\cite{Nico} (see  Appendix C)
confirms  fully the above picture
(Fig. 13a )  as
long as the  instanton anti-instanton distance is large enough:  $R/\rho  >
\sqrt{4/3}  \sim 1.1547... .$

  ($"\sqrt{4/3} \, " $ is only a fit to our numerical
data. However, as can be easily seen from Eq.(\ref{1_7}) and Eq.(\ref{c0}),
 the maximum of the
variation of the Chern Simons number ($C(0)$ ) reaches $1/2$  precisely at
$R/\rho =\sqrt{4/3}$  ($z=3$).   This  makes one
suspect that the critical separation is exactly $\sqrt{4/3}.$ )

On the other hand,  for      $R/\rho $ less than the critical
value $ \sqrt{4/3} $,
  no level crossing is found to occur.
 The
situation is  shown in Fig. 13b.
If  we  identify the level crossing with the (anomalous)  chiral fermion
 generation  or annihilation (as is  the case
in the single  instanton or
anti-instanton background), then we arrive at the conclusion that
the valley field   with   strongly overlapping  instanton anti-instanton
configurations   ($R/\rho  <
\sqrt{4/3} $ ) has nothing to do with anomaly: it is a
purely perturbative background.   The implication of this on the question
of $\Bvio$ cross section in the standard electroweak  theory will be
discussed in the next subsection.

Furthermore, it was found numerically that  at   a  crossing point $t^*$
(the value of $t$ such that a normalizable solution exists),   the
relation
\beq    C(t^*) =  \Delta {\cal N}_{CS}  = {1 \over 2}\eeq
always holds.   This  result  is, after all, very natural since  $
\Delta {\cal N}_{CS}  = {1 \over 2} $   corresponds  precisely to  gauge
fields sitting on top of the hill between the two adjacent
vacua.

\bigskip
To conclude,  the results described in subsection 5.3 and 5.4 imply that
the "instanton anti-instanton" valley configuration ceases to be
topologically significant when the parameter $R/\rho $ is equal or
smaller than the critical value, $\sqrt{4/3},$
 or put more intuitively, when the instanton pair "overlaps" substantially.
 But this means that the lower portion of the valley has not a well-defined
 physical meaning:  it is just  a
 sort of perturbative field, to be  considered together with  generic
 fluctuations around $A_\mu
 =0.$

 Such a result is  not really surprising. In non-Abelian gauge
 theories the perturbative series  is divergent  and believed not to be even
  Borel summable\cite{thooft}.
 In other words, the perturbative series  alone does not
  define  the theory. This is no reason for despair however: we {\it know} that
  in these theories
  there are also physical, non-perturbative effects to be taken into
  account. It is to be   expected that only the {\it sum} of perturbative and
  non-perturbative contributions have a well-defined
 physical meaning, not each of
them  separately. The
systematic  treatment of such a summation and a better
  definition of the theory, let alone the solution to
 the unsolved problem of renormalons,
  are  however not  known at present.

\subsection {High-energy  $\Bvio$ electroweak scattering}
Let us come back to the physics  of $\Bvio$ cross sections at
high energies.    Khoze and Ringwald\cite{kr91}
  have made  a simple model calculation
via optical theorem, see Section 1.4, in
which the valley solution Eq.(\ref{valley}) of the pure $SU(2)$
 Yang-Mills theory is  used in
conjunction with the Higgs contribution in the action,  $ - \pi \rho ^2
 v^2.$   By  putting simply a product of the  standard
  zero modes  for the
external particles (they actually  considered a forward elastic amplitude
with external gauge bosons,  hence these zero modes are replaced by
instanton  or anti-instanton solution).    By performing the integrations
over the collective coordinates with the saddle point method,  they found
that the cross section grew  with energy and at an energy of the order of
the sphaleron mass,
 $x = x_{K\!R}= 8 \sqrt{3/5}$
the exponential t'Hooft suppression was overcome
completely. See Fig. 4a.

Although it is a toy model calculation (the valley equation used there
does not take into account the Higgs coupling to the gauge fields),  it
is worthwhile to reflect  upon its meaning.
 Since it uses the optical theorem
it in principle takes into account the unitarity constraints:  how did
it manage to evade  the "half-suppression" result  mentioned earlier?

The answer appears to be that  a  naive treatment of the external
particles has led  one  astray.   Indeed, the saddle point values of
$R$ and $\rho $  (which depend on the energy)  are such that at $x_{K\!R}$,
$\rho =0;\, R/\rho  =0.$  See Fig. 4b.
  The "valley"  field  with these parameters
 is simply a
perturbative vacuum, $A_\mu =0$.  No wonder  the cross section is
 unsuppressed!

Moreover, the analysis reviewed in the two preceding subsections strongly
suggests that the transition to the purely perturbative  regime occurs
actually much earlier, when the parameter $R/\rho$ reaches $\sqrt{4/3}$.
Translated  into the value of the valley action,  it means that  only the
 "higher" portion of the valley with
\beq   S> {16\pi ^2\over g^2}(0.5960...)    \eeq
has anything to do with  the  $\Bvio$ cross section.    Taking into
account all the other terms,  we conclude that
in this toy-model calculation,
the  $\Bvio$ cross section  is always suppressed at least by
\beq   \sigma_{\Bvio}  \le  e^{ - { 4\pi \over \alpha }(0.3185...)}.   \eeq

\bigskip
Generalizing the lesson learned above,  one can
give an argument for the "half suppression" result  which does
neither depend  on the particular valley trajectory  Eq.(\ref{valley})
nor on its unjustified  use, but only on a number of  general assumptions.
The assumptions needed are the following.

\noindent (1) There is a semi-classical
Euclidean, real background $A_\mu $  (or more properly, an ensemble of
them as in  the valley trajectory) which dominates the four point function, to
be continued and LSZ-amputated to yield the "anomalous" imaginary part
of the elastic amplitude.

 Although this is just an  assumption - quite a  nontrivial one - ,
 and it
might
 sound rather naive, after  all those  discussions about the complex saddle
point fields reviewed in Section 2, there is a clear advantage here. As one
uses the optical theorem here, all final particles are implicitly already
summed over (unitarity!);
the only
effect of the external particles are four fermion
 (or bosonic, if one
prefers to study purely bosonic amplitude) fields, appearing as
pre-exponential factors in the functional integral. It is much less
likely here (as compared to the calculation of the {\it production amplitude},
to be squared and summed over the final states) that the external
particles affect the relevant gauge background.
We recall also that in the approach by Diakonov and Petrov\cite{Diakonov93}
(see Section 3),
 the relevant fields are real and
Euclidean (although singular).

\noindent (2) This gauge background, to
 be able to give rise to the
"anomalous" imaginary part, must have sufficiently large variations of the
Chern-Simons number:  more precisely we assume that $C(t)$ crosses the
value $1/2$ at least twice (once upwards and once downwards, as required by
$Q=0$).  Namely, we
suppose that
\bea
  {\cal N}_{CS}(t_1)-{\cal N}_{CS}(-\infty )  = {1 \over 2},\non \\
  {\cal N}_{CS}(t_2)-{\cal N}_{CS}(\infty )  = {1 \over 2},  \label{fame}  \eea
($t_1 \le t_2,$)
as a (necessary) condition for the  fermion level crossings (to and back,
as in the valley with $R/\rho > \sqrt{4/3} $ ) to occur.
The action of such a gauge background is easily shown to be at least as
large as the single instanton action, $2\pi /\alpha. $  Indeed, since
$\Tr (F_{\mu \nu } \pm {\tilde F}_{\mu \nu })^2 \ge 0, $   for each $x$, it
follows from Eq.(\ref{fame}) and Eq.(\ref{4_1}) that
\bea
 S &=& {1\over 2} \int d^4x\,\Tr F_{\mu \nu }^2 \non \\
 &\ge&
{1 \over 2} \int_{-\infty}^{t_1}\! dt  \int d^3x\, \Tr F_{\mu \nu }^2
 + {1 \over 2} \int_{t_2}^{\infty}\! dt \int d^3x\, \Tr F_{\mu \nu }^2
 \non \\
&\ge&
  -{1 \over 2} \int_{-\infty}^{t_1} \! dt \int d^3x\,
\Tr F_{\mu \nu }{\tilde F}_{\mu \nu }
 + {1 \over 2} \int_{t_2}^{\infty} \! dt \int d^3x\, \Tr F_{\mu \nu }
{\tilde F}_{\mu \nu } =
{2\pi \over \alpha }.
\label{action}\eea
There are also the exponent depending on the initial  energy (coming
from the external particles through the  LSZ procedure),
 the part of the action due to
the Higgs particle,
 as well as an integration over the collective coordinates. The saddle point
equation  however normally has   a solution in which each term of the
exponent is of the same order of magnitude.  This leads to the anomalous
cross section, suppressed by something like the square root of the
t'Hooft factor, in view of Eq.(\ref{action}).

This argument is admittedly not  very rigorous, and should be
regarded at best as a tentative one.    Nevertheless, it concisely
takes into account the two main ingredients of the whole problem,
 {\it unitarity} and {\it
anomaly} in an essential manner.   Further efforts  are   welcome to
see  whether it
 can be made  more rigorous.

\subsection {Note on anomaly for massive fermions}

All calculations described up to now  are  done for massless fermions,
for simplicity.  In the actual world all known fermions (except perhaps for
some or all of the
neutrinos) are massive.   On the one hand, one believes that the physics
at high energies, $E \gg m_f$, ($m_f$ standing for  a generic fermion mass)
should be the same as that with  $m_f =0.$   On the other hand it is not
at all obvious that  anomalous processes such as Eq.(\ref{bvprocess}) do
occur
with  massive fermions. Consider for instance the description of chiral
anomaly in terms of the level crossing (recalled at the beginning of
subsection 5.4).  The smallest (in the absolute value) eigenvalue of the
Hamiltonian which  crosses zero once - in the case of an instanton
background - has the magnitude which is really very close to zero:
$E_1(\pm \infty)  = O({1 \over L}),$  if $L$ denotes the linear size of
the spacetime volume in which our system is put.  If the fermion is
massive, the positive and negative energy levels are separated by a
finite gap $2m_f$; how  an infinitesimal "level crossing" (from
$E_1(- \infty)$ to $E_1(+ \infty)$ ) can cause a jump from  e.g, a  negative
sea level  to a positive one?

The resolution of this  paradox (due to Krasnikov et al.\cite{krasn}
and  Anselm et al.\cite{ansel}; see also Axenides et al.\cite{axeni})
 hinges upon  the very way fermions
get mass in the  Weinberg-Salam theory (the Higgs mechanism).
A fermion mass originates  from the Yukawa interaction,
\beq    L^{(up)}_{Yukawa} = g \,\epsilon
 {\bar \psi }_L \phi  \psi _R  + h.c.  \eeq
if  the Higgs field has a constant part, $\phi  = \pmatrix{ 0 \cr  v/\sqrt{2}
}
+ \ldots$

  Now  the (constrained) instanton responsible for
anomalous process Eq.(\ref{bvprocess}) behaves as  Eq.(\ref{instanton})
near the instanton center.  From Eq.(\ref{instanton}) it follows that
 the fermion mass term vanishes  at the instanton center, i.e, precisely
 where the eigenvalue of the Dirac Hamiltonian changes sign. For this
 reason  the massive fermion experiences the level crossing just as a
massless one.  "Massive zero modes"  have been constructed, and an
analogous paradox is solved in the case of the global
SU(2) anomaly.\cite{axeni}




\section{Conclusion}

A unified physical picture seems to emerge from various different types of
analyses reviewed here. A single instanton in the Euclidean formulation
describes the vacuum-to-vacuum tunnelling and, as such, is not directly
related to the $\Bvio$ electroweak transition in TeV-energy scattering
processes.  The overlap of a high-energy two-particle state with  most
likely  multiparticle states of the same energy and quantum numbers
{\it within
the same well},
enters as a multiplicative factor in the amplitude.
 The rest of the amplitude describes
the tunnelling between states of
 adjacent "wells", whose rate rapidly grows with
energy, and  loses the exponential suppression altogether
 as the  energy exceeds the barrier height, the sphaleron energy.

 The first factor, which describes the mismatch between the "most
favorable"
 state (for the purpose of  barrier penetration) and the initial
  two-particle state,   gets  instead strongly suppressed as the
 energy grows, as has been elegantly illustrated in several quantum
 mechanical
 analogue problems, reported in Section 3.
  It is indeed an analogue of the Landau's
 semi-classical  matrix elements\cite{landau}.

  It has at high energies
the same
 semi-classical form, $\exp - {1\over \alpha }$ as the familiar
 tunnelling factor, although it describes a transition within the same well.
 Ultimately, this factor seems to be responsible for the exponential
 suppression of the $\Bvio$ cross section at high energies found in
 different analyses.  The same factor is found also in the unitarized
 multi-instanton approximation (compare Eq.(\ref{resummed}) with
 Eq.(\ref{nosupp})).

 In  the instanton description, the
 energetic initial particles must first convert themselves into
 $O({1\over \alpha })$ gauge-bosons,  Higgs, plus the original fermions,
 so that these  particles, having small energies, can be absorbed by the
 instanton without any form-factor suppression.  The price paid for the
 conversion, $ \sim  (\alpha )^{1/\alpha } $, gives rise to the mismatch
 factor\cite{mu90}.

The same Landau   factor  seems to be
 at the origin of the qualitative difference between the
$\Bvio$ transition at high temperature or at high fermion density where
the  transition rate  becomes eventually unsuppressed,
 and the  high energy scattering in which baryon number violation is most
 likely to remain unobservable.

All in all, consistency of the result (at least "half-suppression"
 of   $\Bvio$ cross sections)  found from a  variety of different
kinds of analysis (Section 3, Section 4 and Section 5)
 seems to be quite remarkable.  This, together with its physical
understanding (in terms of the Landau factor as well as  in terms of
 the general unitarity bound),  provides us with  certain degree
 of confidence in the  result.

 To be scrupulous, the final outcome of  new, semi-classical approaches
 (reviewed in Sections  2 and 3)
 is not known as yet.
 However, {\it  if\,}  these
analyses  eventually indicated an unsuppressed $\Bvio$ cross sections
at the sphaleron energy, it would come  as a surprise.  It would  be a
new (and very interesting) phenomenon, which has however very little to
do with the original sphaleron or
instanton calculations, and  for which no hints exist for the moment.
But, of course,  who knows?

Barring such a possibility, then, have we made just "MUCH ADO FOR NOTHING",
 after all?

  There
are reasons to believe that this is too pessimistic a viewpoint. The original
suggestion\cite{ring90}\co \cite{esp90} that the instanton cross section is
corrected by an exponentially growing factor at {\it low energies\,} is now
well established. If the enhancement is so as to convert the 't Hooft
factor into something close to its square root  (an optimist would
call it
"half-enhancement"),
\beq e^{ - {4 \pi \over \alpha }} \longrightarrow  e^{ -{2 \pi \over
\alpha }},
\eeq
this will be insignificant in the electroweak theory, but might  be rather
important in the low-energy instanton physics in QCD.   In spite of the
generally accepted belief that the instantons play an  important role in the
physics of low-lying pseudoscalar mesons (for instance, in the solution
of the "U(1)
problem"\cite{u1}),
  quantitative understanding of the role of instantons in these
 systems has not yet  been achieved.  Any improved understanding
 of  physics
related to  instantons should be helpful.   The same can be said,
 with an even stronger emphasis,
with
regard to the instanton liquid
model of the QCD vacuum\cite{liquid}.

Finally,
 all these issues are deeply related to the large-order  breakdown
of perturbation theory and connected problems\cite{zinn},
especially  in  QCD\cite{bac92}\co\cite{lar}.  In our opinion  no truly
significant step
forward
has been made  after the  't Hooft's work\cite{thooft} in this regard.  It is
hoped that somewhat  improved understanding of  instanton physics
achieved through the study of baryon number violation in high-energy
electroweak interactions as  reviewed here, will help clarifying  these
issues as well in a near future.

\appendix {~Saddle-point evaluation of multi instanton sizes}

 The integral over $t_i$, in Eq.(\ref{sono}-\ref{stufo}) can be
easily evaluated since it is factorized: it leads to the
substitution,
\beq
 -iEt_i + (\Cost)\rho_i^2 \rho_{i+1}^2/{t_i^4}
 \longrightarrow 5\Bigl({\Cost \rho_i^2 \rho_{i+1}^2 \over 4^4 E^4}
 \Bigr)^{1/5}
\qquad (i=1,2, ... 2n-1)
\eeq
in the exponent, Eq.(\ref{Wmulti}).
The final integrations over the instanton sizes have the form,
\beq
\int \int ...\int \prod_{i=1}^{2n}d\rho_i \,({\rm powers ~ of} \;
 \rho {\rm 's})
\, \exp (\bar W)
\eeq
where
\beq
\bar W \equiv -\pi^2 v^2 \sum_{i=1}^{2n} \rho^2
+ 5\sum_{i=1}^{2n-1}\Bigl({\Cost \rho_i^2 \rho_{i+1}^2
\over 4^4 E^4 }\Bigr)^{1/5}.
\eeq
For the purpose of performing the above integration by the saddle point
method one can introduce the dimensionless variables $z_i$ instead of
$\rho$'s,
\beq
\rho_i = \bigl((\Cost)E^4 \pi^{10} v^{10} \bigr)^{-1/6} \, z_i^{5/2},
\eeq
in terms of which the exponent becomes
\beq
\bar W = \bigl({96 E^4 \over \pi^2 g^2 v^4}\bigr) Y,
\label{saigo}\eeq
\beq
 Y \equiv - \sum_{i=1}^{2n} z_i^5 + \bigl({5\over 4^{4/5}}) \sum_{i=1}
^{2n-1} z_i z_{i+1}.
\label{Y}\eeq
  The saddle point equations for $z_i$, though simple, do not appear to be
solvable exactly for generic value of $n$. However for $n$ sufficiently
large (in practice it means $n \ge 6 $: see below) the solution can be
found easily numerically.  One finds
\beq
z_i= z_{2n-i+1}, \quad (1=1,2,...n),
\eeq
and
$$
   z_n \simeq z_*=2^{-1/5} = 0.87055056... ;\quad n\ge 6,
$$
$$
  z_1=0.72808;\qquad
  z_2=0.85185;\qquad
  z_3=0.86817;\qquad
$$
\beq
  z_4=0.87025;\qquad
  z_5=0.87051;\qquad
  z_6=0.87055 \simeq z_*
\eeq
Inserting the above solution in Eq.(\ref{Y}) one finds
\beq
Y\simeq {3\over 2}n - 1.06386.
\label{Yres}\eeq
The fact
that the first and last few $z_i$'s differ from $z_*$ -  an end
point effect - is responsible for the
second term of Eq.(\ref{Yres})) and is crucial for
 the subsequent discussions.
Substituting Eq.(\ref{Yres}) into Eq.(\ref{saigo}) one finds Eq.(\ref{A2n}).


\appendix {~Proof of absence of  the fermion zero modes in the valley}

The proof of the absence of the zero modes of the Dirac operator
${\cal D} \equiv i \gamma _{\mu} (\de_\mu -ig A_\mu)$
in the valley background is straightforward.
The gauge field is given by  Eq.(\ref{valley}) which can be  written as
\beq
A_\mu =-{i\over g} (\si _\mu \sb _\nu -\delta_{\mu \nu})
\, F_\nu \equiv \bar{\eta}_{\mu\nu}^a F_\nu \si ^a
\eeq
\beq\label{effemu}
F_\mu={1\over 2}\,\de_\mu\log L(x)\;\;\;\;\;\;
L(x) \equiv {(x-x_a)^2+\rho^2\over (x-x_i)^2+\rho^2}\,(x-x_i+y)^2.
\eeq
We consider the decomposition:
\beq
 {\cal D}=\pmatrix {0 & \dminus\cr
                 \dplus &0}.
\eeq
Because the  valley is a superposition of an instanton
 (with a righthanded zero mode)
 and an anti-istanton field, (with a lefthanded  one), the only physical
possibility (to be ruled out) is that {\it both} $\dplus$ and $\dminus$ have
a single, non degenerate, zero mode.

Using the explicit form of $\eta$\cite{thoft76}), one has
\beq
\dplus\equiv -\de_t +i\vec{\si}\cdot\vec{\nabla} +i\vec{\tau}\cdot\vec{F}
-\vec{\si}\wedge \vec{\tau}\cdot\vec{F}-F_0 \vec{\si}\cdot\vec{\tau}
\eeq
\beq
\dminus\equiv +\de_t +i\vec{\si}\cdot\vec{\nabla} -i\vec{\tau}\cdot\vec{F}
-\vec{\si}\wedge \vec{\tau}\cdot\vec{F}-F_0 \vec{\si}\cdot\vec{\tau}
\eeq
 where $x_\mu\equiv(t,\vec{r})$,
 \beq\label{noncovariant}
\vec{F}=\vec{r}f(r,t);\;\;\; F_0=F_0(r,t)
\eeq
$r=\sqrt{{\vec{r}}^2}$,  $\vec{\si} $ are spin operators, and
$\vec{\tau } $ are isospin one.

Since $\dirac$ commutes
with the three dimensional
total angular momentum operator
$\vec{J}=-i\vec{r}\wedge{\vec \nabla}+\vec{\si}+\vec{\tau},$ the requirement
of having a nondegenerate zero mode  is satisfied if $j=0$.

Let us concentrate on the left-handed mode, associated with $\dplus$
below.
The most general function with $j=0$ can  be written
accordingly\cite{guida}  as:
\beq\label{singlet}
\eta_{j\alpha}=
(\si_2)_{j\alpha}S(r,t)-i(\vec{\si}\si_2)_{j\alpha}\cdot\vec{r}\,T(r,t)
\eeq
The equation $\dplus \eta =0$ is equivalent to:
\beq
 \matrix{\de_t \st - h\de_r \tt&=0;\qquad
-\de_r \st - h\de_t \tt&=0,}
\eeq
where
 $h=h(r,t)={L(x)^2\over r^2}$ and two new functions $\tt$ and $\st$,
 \beq
T={1\over r^3}{L}^{1\over 2} \tilde{T};
\qquad S=L^{-{3\over 2}} \tilde{S},
\eeq
 have been introduced.
The consistency condition
following from  the above  is equivalent
to $\nabla (h \nabla \tt)=0$ (here $\nabla\equiv (\de_r,\de_t)$).
Furthermore
the normalization condition reads
$$
\int d^4x\, r^2|T|^2 =
\int_{-\infi}^{+\infi}dt \int_{0}^{+\infi}dr\,
4\pi {L\over r^2}\tt^2 <\infi;
$$
 \beq
 \int d^4x \,|S|^2 =
\int_{-\infi}^{+\infi}dt \int_{0}^{+\infi}dr\,
  4\pi r^2 {L}^{-3}\st^2 <\infi.
 \label{normali} \eeq
Note that
with our choices of $x_i=(-{R\over 2},0,0,0)$ and $x_a=({R\over 2},0,0,0)$,
 $L(r,t)\neq 0$ almost everywhere on the line $r=0$,
so that a necessary
condition  for normalization is $\tt (0,t)=0$.

One is thus led  to the boundary value problem:
\beq
 \matrix {\nabla (h \nabla\tt )=0,&\;\; \hbox{\rm on }\Omega;\cr
                 \tt =0,&\;\; \hbox{\rm on }\de\Omega;\cr
	         \tt\in H^2(\Omega ),&\;\;\tt\in C^0(\bar{\Omega}),}
	         \eeq
where $\Omega \equiv\{(r,t)\in {\bf R}^2/r>0\}$.

The operator $\,\nabla^2+\nabla h\nabla \,$ is an elliptic differential
operator
 with coefficients analytic in $\Omega$. At this point one can use standard
theorems  to prove that
 $\tt$ cannot have global extrema
 on $\Omega$.  Using the continuity
on $\bar{\Omega}$, the boundary conditions Eq.(\ref{normali})
and the condition
$\tt \rightarrow 0, \, r,t \rightarrow \infty, $
it is easy to prove that $\tt =0$.

It follows that $\st $ must be a constant, i.e,
$$ S = {\rm const.} \, L(x)^{-{3\over 2}}. $$
 The normalization condition however
forces $S =0$.  {\it Q.E.D.}
\smallskip

Having  disposed of the zero modes,
 one might then ask if the zero mode of $\dplus $ in the antistanton
field, and that of $\dminus $ in the istanton background, combine somehow
 in the valley field so as  to form  a  pair of non zero
 eigenvalue $ \pm \lambda$
of $\dirac$.  A theorem by Ikebe and Uchiyama\cite{KYOTO}
 however excludes this
possibility.

\appendix {~Level-crossing in the valley}

The Dirac Hamiltonian $\H$ is defined by:
$
i\gamma _\mu D_\mu=i\gamma_0 (D_0 + \H).
$

Decomposing
$$
\cal{H}= \pmatrix {H_{+} & 0\cr
                 0 &H_{-}},$$
and passing to a noncovariant formalism,
 one  finds:
\beq
 H_{+}=-H_{-}=H,
\label{condition}\eeq
 with
\beq
H(t) =+i\vec{\si}\cdot\vec{\nabla}
-\vec{\si}\wedge \vec{\tau}\cdot\vec{F}-F_0 \vec{\si}\cdot\vec{\tau}
\label{hamiltonian}\eeq
 where
the  notation is the same as in Appendix B, Eq.(\ref{noncovariant}).

To simplify  the problem, one can just look for zero modes of
${\cal H } (t; R)$
without studying the detailed
 behavior of levels in terms of parameters $t , R$.

The  condition (\ref{condition})
implies that zero modes always appear in pairs of opposite  chirality.
We assume that for each chirality  zero modes are non degenerate, so that
we must look for singlets of the total spatial
angular momentum (which is a symmetry
of $\H$).
The most general form of
the singlet $\eta$ is given in (\ref{singlet}).

The equation $H\eta =0$ then reads:
\bea
3 F_0 S-3T-\vec{r}\vec{\nabla}T+2 \vec{F} \vec{r} T&=&0,   \non \\
-F_0 r^2 T +\vec{r}\vec{\nabla} S+2\vec{F} S&=&0.
\eea

Making the substitution
\beq
S={1\over L}\st, \qquad
T={L\over r^3} \tt,
\eeq
 we get a  simplified system:
\beq \label{system}
\tt '(r;t)=3F_0 {r^2 \over L^{2}} \st ; \qquad
\st '(r;t)=F_0 {L^2\over r^2} \tt
,\eeq
where the prime means differentiation with respect to $r$ and the
time $t$ appears as a
 parameter.

Normalizability of the solution implies:
\beq \label{normalization}
\infi >\int_0^{\infi}d\!r r^2  |S|^2=\int_0^{\infi}d\!r {r^2\over L^2}
|\st|^2 ;\qquad
\infi >\int_0^{\infi}d\!r r^4 |T|^2=\int_0^{\infi}d\!r {L^2\over r^2}
|\tt|^2.\eeq

We wish  to know for which range of the parameter $R$ and for which values
of $t$ - if any - (the other
parameter $\rho$ just fixes the scale) the system Eq.(\ref{system}) has
a normalizable solution.
Normalizability
enforces the following initial condition (at a given $t$ )
\bea
&&\st (0)=1;\;\; \tt (0)=0;\;\;\hbox{\rm if}\;\; t\neq y+{R\over 2}
\label{initial1}\\
&&\st (0)=0;\;\; \tt (0)=1;\;\;\hbox{\rm if}\;\; t= y+{R\over 2},
 \label{initial2}
\eea
where $y$ is defined (\ref{1_7}).

The  problem turns out to be
 too hard  to be
 treated analytically: it suffices to note that the system is equivalent to
a second order linear equation for $\tt$  with $12$ regular fuchsian points!
In passing we recall\cite{Kiskis}
 that for the case of a pure anti-instanton
(or an instanton)
a normalizable solution can be found analytically.
Lacking a better method  a numerical method  was used  to solve
 Eq.(\ref{system}).    The strategy adopted is as follows.

\noindent(i)  First solve the system by a power series in $r$  at
$r \simeq 0 $  and $ r  \to  \infty$.    It turns out that in  each
region  one
and only one of the two independent solutions is compatible with
the normalizability.

\noindent(ii)  Choose, for a given $t$ and $R/\rho $,  the normalizable
solution for $\tt, \st$  near $t =0$.    Make them evolve according to
Eq.(\ref{system}) up to a  very large but fixed value of $r$,  $r_\Lambda $.

\noindent(iii)   Plot $\tt(r_\Lambda )$  as a function of $t$,  for a fixed
$R/\rho $.   See whether $\tt(r_\Lambda )$   crosses  zero as $t$ is
varied.   If it does at some $t$,  it means  by continuity
that for $t$ very close to
that value there is a  normalizable solution.   The power series solution of
the coupled equation assures that whenever  $\tt$ is normalizable, so is
$\st$.

\noindent (iv) Repeat the same procedure  for different values of
$R/\rho $.

Note that this method is somewhat  similar  to  that used in the proof of the
so-called oscillation theorem in  one-dimensional
quantum
 mechanics\cite{messiyah}.
   It is also reminiscent of  the "shooting method" used
by mathematicians for solving certain differential equations.

The result cited
 earlier (Fig. 13a   and Fig. 13b     )   was obtained   this way:
for all values of $R/\rho $ above a critical value,
\beq{R_c\over \rho} \simeq 1.15470....\simeq \sqrt{{4\over 3}},\eeq
there are two values of $t$ for which a normalizable solution of
$H \eta  =0 $ exists.    Below the critical separation, no level crossing
occurs.

\nonumsection{ Acknowledgements}

Part of this work was done during a visit of the authors  to
 CERN. We thank the members of its Theory Division  for
hospitality.
We  thank A. Ringwald and H. Suzuki for useful discussions. One of the authors
  (K.K.)
 wishes  to thank the Physics Department
of  University of Pisa for hospitality.



\nonumsection{ References}

\nonumsection{ Figure captions}

\begin{description}
\item{Fig 1.} Graphs representing:
 a) the leading approximation, b) a soft-soft correction,
c) a hard-soft corrections, d) a hard-hard correction,
 in a theory with fermions (in an instanton background).
\item{Fig 2.} The leading contribution to the exponent
in the  $R$-term approach:
double residue of $W$ classical field integrated over the phase space.
\item{Fig 3.}
 The action of the  instanton anti-instanton valley (in pure gauge theory),
plotted in terms of $R/\rho$.
\item{Fig 4.} a) The saddle point exponent, b) the saddle point ratio $R/\rho$,
versus
 the parameter $x=E/\sp$ in the Khoze-Ringwald model.
\item{Fig 5.} The lowest hard-hard correction in a pure gauge theory.
\item{Fig 6.} Example of "squared tree" multiloop graph
     in the instanton background (blobs means tree graphs).
\item{Fig 7.} The relevant complex-time path in the multiparticle approach.
\item{Fig 8.} The behavior of the singular solution
used in the W.K.B. approach ($B$ and $ C$ are to be intended located
at $-\infty$).
\item{Fig 9.} The contours
$\gamma _1$ and $\gamma_2$
 in the complex $q$ plane, used in the W.K.B. approach.
\item{Fig 10.} A multiinstanton chain. The solid lines indicate  exchange
 of many bosons, the lines with an arrow representing  fermions. For simplicity
of  drawing $N_F$ is taken to be $2$ here.
\item{Fig 11.} $C(x_4)$ versus $x_4/\rho$ for ${R\over \rho} =10$
(outmost curve),
${R\over \rho} =5$, ${R\over \rho} =2$ (middle), ${R\over \rho} =1$ and
${R\over \rho} =0.5$ (innermost curve).
\item{Fig 12.} The maximum of $C(x_4),$   $C(0),$
as a function of ${R\over \rho}$.
\item{Fig 13.} Schematic behavior of the level crossing for the
Dirac Hamiltonian
in the valley background,  a) for $R> R_c$, b) for $R<R_c$.

\end{description}
\end{document}